\documentclass[twocolumn,superscriptaddress,showpacs,preprintnumbers,showkeys,amsmath,amssymb,aps,prb,floatfix]{revtex4-1}
\usepackage{amsmath}
\usepackage{xcolor}
\usepackage{graphicx}
\usepackage{bm}
\usepackage[colorlinks=true, citecolor=blue, urlcolor=blue, linkcolor=blue]{hyperref}
\usepackage[normalem]{ulem}
\usepackage{soul}
\usepackage{gensymb}
\usepackage{siunitx}

\newcommand{\Emax}{$\epsilon^{\text{spin}}_{\text{max}}$}
\newcommand{\CCOC}{Ca$_2$CuO$_2$Cl$_2$}
\newcommand{\SCOC}{Sr$_2$CuO$_2$Cl$_2$}
\newcommand{\NaCCOC}{Na$_x$Ca$_{2-x}$CuO$_2$Cl$_2$}

\newcommand{\LCO}{La$_{2}$CuO$_4$}
\newcommand{\HBCO}{HgBa$_{2}$CuO$_4$}
\newcommand{\CCO}{CaCuO$_2$}
\newcommand{\BSCO}{Bi$_{2}$Sr$_2$CuO$_6$}
\newcommand{\BSCCO}{Bi$_{2}$Sr$_2$CaCu$_2$O$_8$}
\newcommand{\BYSCO}{Bi$_{2}$Sr$_{2}$YCu$_{2}$O$_8$}

\newcommand{\BiTTOT}{Bi$_{2}$Sr$_2$CaCu$_2$O$_8$}

\newcommand{\LaTOF}{La$_{2-x}$M$_x$CuO$_4$ (M=Ba,Sr)}

\newcommand{\p}{^{\prime}}
\newcommand{\pp}{^{\prime\prime}}

\begin{document}
	
	\title{Paramagnon dispersion and damping in doped Na$_{x}$Ca$_{2-x}$CuO$_2$Cl$_2$}
	
	\author{Blair W. Lebert}
	\thanks{These two authors contributed equally.}
	\affiliation{IMPMC, Sorbonne Université, CNRS, MNHN, 4 place Jussieu, 75252 Paris, France}
	\altaffiliation{Present address: Department of Physics, University of Toronto, Toronto, Ontario, M5S 1A7, Canada}
	\affiliation{Synchrotron SOLEIL, L'Orme des Merisiers, Saint-Aubin, 91192 Gif-sur-Yvette Cedex, France}
	\altaffiliation{Present address: Department of Physics, University of Toronto, Toronto, Ontario, M5S 1A7, Canada}
	
	\author{Benjamin Bacq-Labreuil}
	\thanks{These two authors contributed equally.}
	\affiliation{CPHT,  CNRS,  Ecole  Polytechnique,  Institut  Polytechnique  de  Paris,  F-91128  Palaiseau,  France}
	\altaffiliation{Present address: D\'epartement de physique, Regroupement qu\'eb\'ecois sur les mat\'eriaux de pointe \& Institut quantique Universit\'e de Sherbrooke, 2500 Boul. Universit\'e, Sherbrooke, Qu\'ebec J1K2R1, Canada}
	
	\author{Mark P. M. Dean} 
	\affiliation{Condensed Matter Physics and Materials Science Department, Brookhaven National Laboratory, Upton, New York 11973, USA}
	
	\author{Kari Ruotsalainen}
	\author{Alessandro Nicolaou} 
	\affiliation{Synchrotron SOLEIL, L'Orme des Merisiers, Saint-Aubin, 91192 Gif-sur-Yvette Cedex, France}
	
	\author{Simo Huotari}
	\affiliation{Department of Physics, University of Helsinki, FI-00014 University of Helsinki, Finland}
	
	\author{Ikuya Yamada}
	\affiliation{Department of Materials Science, Graduate School of Engineering, Osaka Prefecture University1-2 Gakuen-cho, Naka-ku, Sakai, Osaka 599-8570, Japan}
	
	\author{Hajime Yamamoto}
	\affiliation{Laboratory for Materials and Structures, Tokyo Institute of Technology, 4259 Nagatsuta, Midori-ku, Yokohama, 226-8503, Japan}
	\altaffiliation[Permanent address: ]{Institute of Multidisciplinary Research for Advanced Materials, Tohoku Univ. (IMRAM)
		Katahira 2-1-1, Aoba-ku, Sendai 980-8577, Japan}
	\author{Masaki Azuma}
	\affiliation{Laboratory for Materials and Structures, Tokyo Institute of Technology, 4259 Nagatsuta, Midori-ku, Yokohama, 226-8503, Japan}
	\affiliation{Kanagawa Institute of Industrial Science and Technology, Ebina 243-0435, Japan}
	
	\author{Nicholas B. Brookes}
	\author{Flora Yakhou}
	\affiliation{European Synchrotron Radiation Facility (ESRF), B.P. 220, F-38043 Grenoble Cedex, France}
	
	\author{Hu Miao} 
	\affiliation{Condensed Matter Physics and Materials Science Department, Brookhaven National Laboratory, Upton, New York 11973, USA}
	\affiliation{Materials Science and Technology Division, Oak Ridge National Laboratory, Oak Ridge, TN, USA.}
	
	\author{David Santos-Cottin}                                               
	\affiliation{Department of Physics, University of Fribourg, 1700 Fribourg, Switzerland}
	
	\author{Benjamin Lenz}
	\email{benjamin.lenz@upmc.fr}
	\affiliation{IMPMC, Sorbonne Université, CNRS, MNHN, 4 place Jussieu, 75252 Paris, France}
	
	\author{Silke Biermann}
	\email{silke.biermann@cpht.polytechnique.fr}
	\affiliation{CPHT,  CNRS,  Ecole  Polytechnique,  Institut  Polytechnique  de  Paris,  F-91128  Palaiseau,  France}
	\affiliation{Coll{\`e}ge  de  France,  11  place  Marcelin  Berthelot,  75005  Paris,  France}
	\affiliation{Department  of  Physics,  Division  of  Mathematical  Physics,Lund  University,  Professorsgatan  1,  22363  Lund,  Sweden}
	\affiliation{European  Theoretical  Spectroscopy  Facility, 91128 Palaiseau, France, Europe}
	
	\author{Matteo d'Astuto}
	\email{matteo.dastuto@neel.cnrs.fr}
	\affiliation{Univ. Grenoble Alpes, CNRS, Grenoble INP, Institut Néel, 38000 Grenoble, France}
	
	\date{\today}
	
	\begin{abstract}
		
		Using Resonant Inelastic X-ray Scattering, we measure the paramagnon dispersion and damping of undoped, antiferromagnetic \CCOC{} as well as doped, superconducting \NaCCOC{}. 
		Our estimation of the spin-exchange parameter and width of the paramagnon peak at the zone boundary $X=(0.5,0)$ confirms that no simple relation can be drawn between these parameters and the critical temperature $T_\mathrm{c}$. 
		Consistently with other cuprate compounds, we show that upon doping there is a slight softening at $(0.25,0)$, but not at the zone boundary $X$.
		In combination with these measurements we perform calculations of the dynamical spin structure factor of the one-band Hubbard model using cluster dynamical mean-field theory. 
		The calculations are in excellent agreement with the experiment in the undoped case, both in terms of energy position and width. 
		While the increase in width is also captured upon doping, the dynamical spin structure factor shows a sizable softening at $X$, which provides insightful information on the length-scale of the spin fluctuations in doped cuprates. 
		
	\end{abstract}
	

	\maketitle
	\section{\label{intro}Introduction}
	
\begin{figure}[t]
		\includegraphics[width=1.0\linewidth]{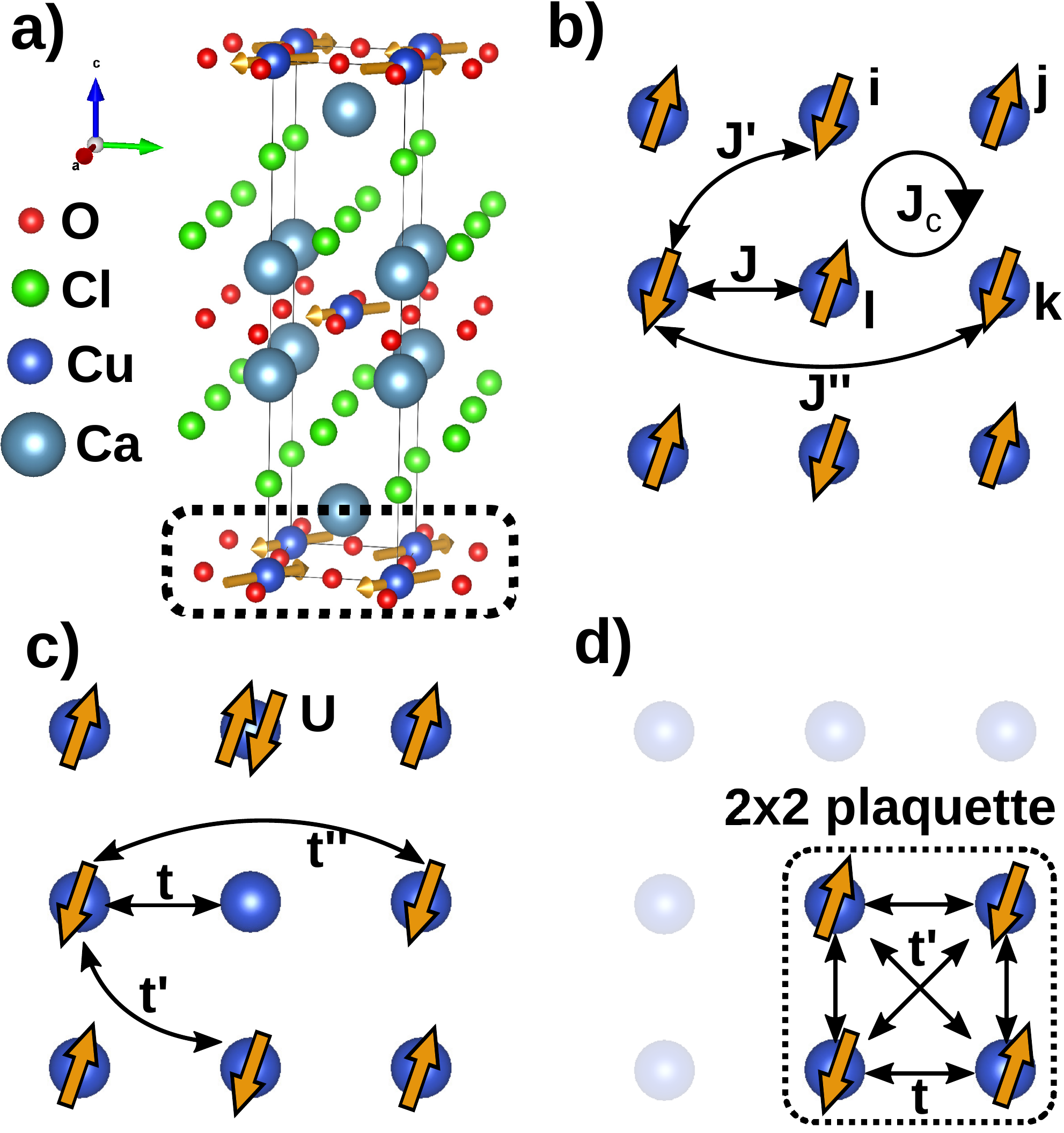}
		\caption{\label{intro-pict} 
			(a) Tetragonal crystal structure of \CCOC{}~\cite{baptiste-ccoc-cif}.  
			The chlorine ions are located in the apical site above and below the copper. 
			Orange arrows indicate one of the possible magnetic structures consistent with neutron diffraction data~\cite{vaknin_prb}. 
			(b) Effective Heisenberg spin model (Eq.~\ref{eq:Hspin}) for the CuO$_2$ layers highlighted in (a) with the dotted framed.
			(c) Effective one-band Hubbard model (Eq.~\ref{eq:Hubbard}), and (d) the $2\times2$ plaquette cluster used for the CDMFT calculation. 
		}
	\end{figure}

	The fate of the spin fluctuations upon doping, as well as their relation to the superconducting critical temperature $T_\mathrm{c}$, are key elements to be clarified for a better understanding of high-$T_\mathrm{c}$ superconductivity in cuprates. 
	Indeed, soon after the discovery of high-temperature superconductivity in cuprates~\cite{Bednorz1986}, the spin-fluctuation exchange mechanism was suggested as a possible pairing glue~\cite{Scalapino1995329,Moriya_2003}.
	Not only would it account for the experimentally observed $d$-wave character of the superconducting gap~\cite{Kirtley-NatPhys2006}, but the spin-exchange parameter $\mathcal{J}$ also seems to provide a correct order of magnitude of the superconducting transition temperature $T_\mathrm{c}$~\cite{tacon-paramag, wang2022}.
	Similarly, the frequency spread of the spin fluctuations appears to be linked with $T_\mathrm{c}$~\cite{Moriya_2003}.
	Recent cutting-edge experimental and theoretical investigations further support the spin-fluctuation exchange scenario~\cite{kowalski2021,omahony2022} through the expected anticorrelation between the superexchange and the charge order gap.
	Nonetheless, spin-fluctuation theory involves uncontrolled approximations and there is no consensus on the role of spin-fluctuation exchange in the pairing mechanism of superconductivity in cuprates.
	In particular, a direct link between $\mathcal{J}$~\cite{wang2022}, or the spin-fluctuation damping, and superconducting $T_\mathrm{c}$ has proven arduous to establish.
	Another crucial concern is the evolution of the spin fluctuations upon doping~\cite{n-mat-dean-lascuo, tacon-paramag, dean-bisco-PRL,guarise-bsccoNC,Dean20153,PhysRevB.98.144507}.
	Far from being suppressed at low doping, the spin fluctuations are persistent even in the overdoped regime~\cite{n-mat-dean-lascuo,meyers-lscoPRB}.
	While for the undoped compounds the spin fluctuations can be accounted for by spin-only Hamiltonians, describing the interplay between charge and spin degrees of freedom
	upon doping remains a challenge \cite{Ayral-Parcollet}.
	Part of the difficulty in solving these problems lies in the lack of knowledge about how the details of the crystal structure of different compounds may influence $\mathcal{J}$~\cite{omahony2022}, as for instance the influence of the number of consecutive CuO$_2$ planes in Hg-based compounds~\cite{mukuda2012}, or the structural distortion in YBa$_2$Cu$_{3}$O$_{7-x}$~\cite{horn1987}, to name a few.
	This prevents to draw a naive causality between the measured $T_{\mathrm{c}}$ and $\mathcal{J}$. 
	From a theoretical point of view, the strongly correlated nature of cuprates restricts the effective low-energy models to idealized 2D systems, usually based on the Hubbard model~\cite{hubbard1963,gutzwiller1963,kanamori1963}.
	\NaCCOC{} (Na-CCOC)~\cite{hiroi,kohsaka-jacs}, formed by Na substitution from the parent compound \CCOC{} (CCOC), is an interesting candidate in this regard.
	Indeed, Na-CCOC displays a simple tetragonal structure $I4/mmm$ as shown in Fig.~\ref{intro-pict}, and no known structural phase transition as a function of doping and temperature. 
	Most interestingly, the replacement of the apical O by Cl atoms confers a strong 2D character to the CuO$_2$ layers.
	Hence Na-CCOC appears as an interesting platform to connect the theoretical models to a real material.

	In this paper we use RIXS to measure undoped antiferromagnetic \CCOC{} as well as doped superconducting \NaCCOC{}.
	We also propose a simple and numerically cheap scheme to determine
	the on-site Coulomb interaction $U$, based on a combination of
	first-principles calculations for the hopping parameters and a
	fit of the magnon dispersion. This trick should be applicable also
	to other cuprate systems and beyond.
	The choice of RIXS as a probe for spin excitation is motivated by the very small volume available for our single crystals, especially the doped ones that are synthesized under high pressure of several GPa\cite{hiroi,kohsaka-jacs, Yamada2005}. 
	Another advantage is the possibility to measure the paramagnon dispersion over a wide range of the Brillouin zone, allowing a reliable estimation of $\mathcal{J}$.
	The dispersion of the parent compound is fit with a spin-model whose parameters are related to the hopping amplitudes $t_{ij}$ and on-site Coulomb interaction $U$ of a one-band Hubbard model.
	By fixing the $t_{ij}$ to the values obtained from an \emph{ab initio} calculation, we are able to estimate $U$ for our theoretical calculations.
	Our fitting procedure provides one-band model parameters consistent with those found in the literature either from fully (computationally costly) \emph{ab initio} methods~\cite{hirayama2018,moree2022}, or from pure fit of the experimental data~\cite{dalla_piazza_rapidB,PhysRevLett.80.4245,ccoc-sw-blair1,dean-bisco-PRL} which can suffer from the non-uniqueness of the fitting parameters.
	This method therefore stands as a cheap alternative to determine $U$ for performing realistic calculations whenever experimental data is available. 
	The estimated $\mathcal{J}=157.1~\si{meV}$ is larger than the ones measured in other compounds~\cite{dalla_piazza_rapidB,PhysRevLett.80.4245,ccoc-sw-blair1,dean-bisco-PRL}, although the $T_\mathrm{c}$ of Na-CCOC is smaller. 
	Moreover, the ratio of the width of the paramagnon to $T_\mathrm{c}$ in Na-CCOC is larger than most other cuprates~\cite{Moriya_2003}.
	Hence, our mesurements confirm that there is no simple relation between either the value of the exchange $\mathcal{J}$ or the frequency spread of the spin fluctuations, and the critical temperature $T_\mathrm{c}$. 
	To support the measurements, we compute the dynamical spin structure factor at the zone boundary $X=(0.5, 0)$ using cluster dynamical mean-field theory. 
	We obtain an excellent agreement in the undoped phase both in terms of width and energy position as compared to the RIXS data, and we also capture the spectral broadening upon doping. 
	Yet, the calculation predicts a softening of the magnon peak at $X$ for the $x=0.1$ doped case, which is not seen in our measurements. 
	Instead, the $x=0.11$ sample yields a dispersion consistent with other cuprate compounds in which no softening is observed at the zone boundary~\cite{Jia2014,chaix2018}. 
	This brings valuable information on the evolution of the spin fluctuation coherence length upon doping.

	\section{\label{sec:methods} Methods}
	
	\subsection{Crystal growth and characterization}
	
	Single crystals of \CCOC{} were grown by the flux method. 
	CaO (99.99\%) and CuCl$_{2}$ (99.99\%) powders were mixed with a molar ratio of 2~:~1 and put into an alumina crucible. 
	The mixed powder was heated to 1053~K for 24 h then grounded again after cooling to room temperature. 
	Then, the \CCOC{} precursor was heated to 1053~K at a ramp rate of 60~K/h and kept at this temperature for 5 h. 
	It is further heated to 1203~K at a ramp rate of 60 K/h and kept at that temperature for 10 h, then cooled down to room temperature at a ramp rate of 60 K/h.
	
	For single-crystal synthesis of Na-doped copper oxychloride, $i.e.$ \NaCCOC{} samples (Na-CCOC), the following precursors were used: CaCO$_{3}$ (99.99\%), CuO (99.99\%), CaCl$_{2}$ (99.99\%), NaClO$_{4}$ (99.9\%) and NaCl (99.99\%). 
	First, we prepared a stoichiometric Ca$_{2}$CuO$_{2}$Cl$_{2}$ powder by a solid state reaction of CaCO$_{3}$, CuO, and CaCl$_{2}$ as described in previous works \cite{hiroi, kohsaka-jacs, Yamada2005}. 
	In an argon-filled dry box, we mixed the resulting Ca$_{2}$CuO$_{2}$Cl$_{2}$ powder with NaClO$_{4}$, NaCl and CuO precursors in a molar ratio of 5~:~1~:~1~:~1. 
	The mixture was then loaded in cylindrical Pt capsules and set in high-pressure cells.
	Since it was shown that the synthesis pressure is related to the Na content \cite{hiroi,kohsaka-jacs}, we compressed the pressure cell between 3.5 to 4 GPa in a cubic anvil type high-pressure apparatus in order to get underdoped Na-CCOC single crystals. 
	The capsules were heated up to 1523 K at a rate of 10 K/min, kept at this temperature for 1 h and then slowly cooled down to 1323 K at a rate of 10–20 K/h. 
	The pressure was released at the end of the heat treatment.
	We obtained \NaCCOC{} single crystals with $x \simeq 0.11$ ($T_{\text{C}}$ $\simeq 14$ K) and $x \simeq 0.16$ ($T_{\text{C}}$ $\simeq 23$ K), checked on the individual crystal using a SQUID magnetometer (Quantum design MPMS\copyright).
	All single crystals were characterized and aligned using x-ray diffraction.
	These measurements yielded unit cell parameters and doping levels in agreement with the literature \cite{hiroi, kohsaka-jacs,baptiste-ccoc-cif}, as well as with the diffraction on the powder grinded after the synthesis to check their quality.

	\subsection{\label{Met:RIXS}Resonant inelastic x-ray scattering}
	Resonant inelastic x-ray scattering (RIXS) measurements were performed on the ERIXS spectrometer \cite{Brookes2018} at the ID32 beamline of the European Synchrotron Radiation Facility (ESRF). 
	To avoid hygroscopic damage of the surface, the samples were cleaved under Ar atmosphere before being loaded into the experimental chamber. 
	The total energy resolution was $\Delta$E $\simeq$ 85~meV Full Width Half Maximum (FWHM) as measured on carbon tape. All data was collected at 22~K.

\begin{figure}[t]
		\includegraphics[]{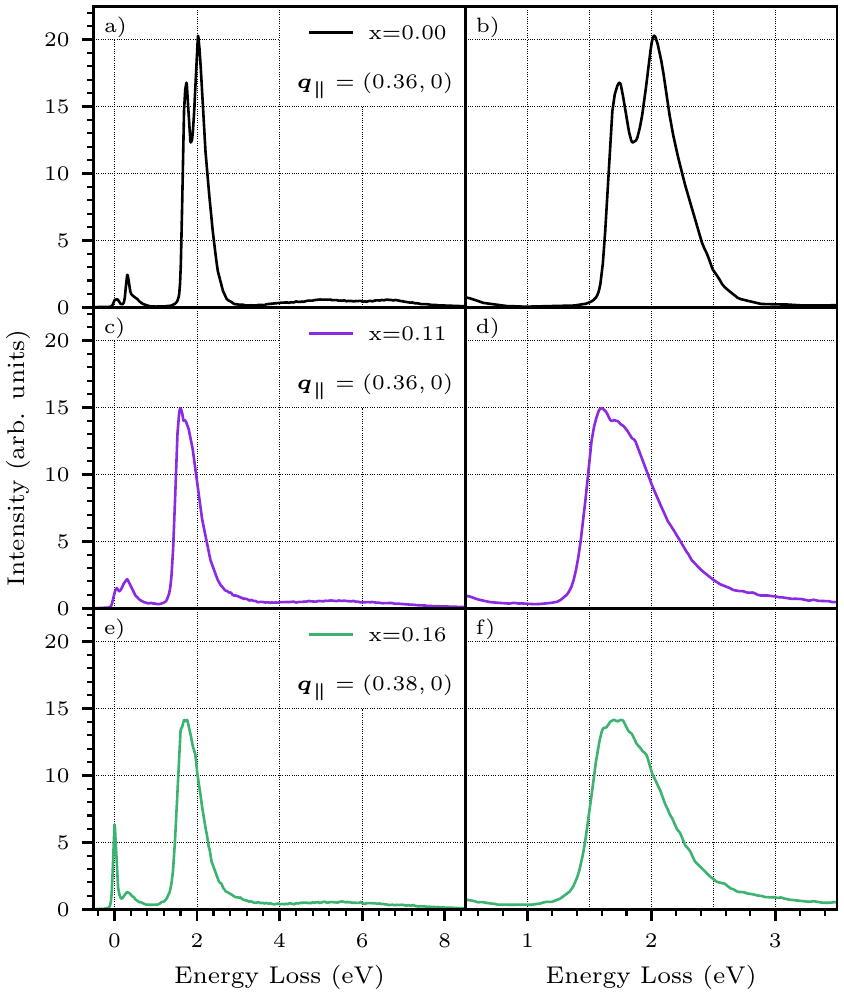}
		\caption{\label{RIXS-example} 
			Cu L$_3$-edge RIXS spectra measured around the in-plane projected wavevector $\bm{q_\parallel}=$ (0.37, 0) for x $=$ 0.00 (a,b), x $=$ 0.11 (c,d), and x $=$ 0.16 (e,f) \NaCCOC{} samples.
			The left column (a,c,e) shows the full energy domain with the quasielastic lines and phonons at lower energy, the magnetic excitations around 0.2--0.4~eV, the $dd$ excitations around 1--3~eV, and the charge-transfer excitations around 4--8~eV. 
			The right column (b,d,f) displays a zoom on the $dd$ excitations to highlight their evolution with doping. 
		}
	\end{figure}

	The incident x-ray energy was tuned to the Cu L$_{3}$ edge ($\simeq$~931~eV) with $\pi$ polarization. 
	The scattered x-rays were measured at a fixed angle of $2\theta = 149.5$\degree{}. 
	The samples were mounted with the $c$ axis corresponding to the azimuthal direction and laying in the scattering plane. 
	The azimuthal angle was rotated to probe along either ($h$, 0, $l$) or ($h$, $h$, $l$), using reduced length units of $(\frac{2\pi}{a},\frac{2\pi}{a},\frac{2\pi}{c})$. 
	However, due to the quasi-2D nature of \NaCCOC{} we only consider the in-plane momentum transfer, i.e. $\bm {q_{\parallel}}$ $=$ ($h$, $k$), along the main symmetry line $k=0$ (with zone boundary $X=(0.5,0)$) and $k=h$ (with zone boundary $M=(0.5,0.5)$). 
	The momentum transfer was varied by rotating the sample along the polar angle and is reported in reciprocal lattice units (r.l.u). 
	We measured all samples in grazing emission geometry, represented here as positive $h$, since it is known to enhance the spin-flip excitations\cite{pol-dep-prb} when coupled with the $\pi$ polarization.
	Representative examples of the full measured spectra are shown in Fig.~\ref{RIXS-example} for $\bm {q_{\parallel}}$~$\simeq$~(0.37,~0).

	\subsection{\label{Met:CDMFT} Dynamical spin structure factor calculations}
	
	In support of the RIXS measurements, we performed calculations of the dynamical spin structure factor, $S(\bm{Q},\omega)$, for the one-band Hubbard model.
	The latter is a minimal model which is believed to describe low-energy properties~\cite{civelli2005,kyung2006,macridin2006} of cuprates, and especially the effects of the strong spin fluctuations as exemplified by the \emph{waterfall} feature~\cite{damascelli2003,ronning2005,macridin2007,manousakis2007,wang2015}.
	Moreover, it also provides an accurate description of the spin fluctuations themselves~\cite{Jia2014,stepanov2018}, and is therefore a reasonable model to study the paramagnon properties of \NaCCOC{}.
	The model is illustrated in Fig.~\ref{intro-pict}(c) and defined as follows:
	\begin{equation}
		\label{eq:Hubbard}
		\begin{split}
			H=&U\sum_i n_{i\uparrow} n_{i\downarrow} - \mu\sum_i n_{i} - t\sum_{\langle ij\rangle,\sigma} c_{i\sigma}^\dagger c_{j\sigma}^{\phantom{\dagger}} \\
			& - t\p\sum_{\langle\langle ij\rangle\rangle,\sigma} c_{i\sigma}^\dagger c_{j\sigma}^{\phantom{\dagger}}
			- t\pp\sum_{\langle\langle\langle ij\rangle\rangle\rangle,\sigma} c_{i\sigma}^\dagger c_{j\sigma}^{\phantom{\dagger}},
		\end{split}
	\end{equation}
	where $U$ is the Hubbard on-site interaction term, $\mu$ the chemical potential, $\langle .\rangle,\langle\langle .\rangle\rangle,\langle\langle\langle .\rangle\rangle\rangle$ denote respectively the nearest-neighbor (NN), next-NN, and next-next-NN associated with the corresponding hopping terms $t=0.425~\si{eV}$, $t\p=-0.076~\si{eV}\simeq-0.18/t$ and $t\pp=0.05~\si{eV}\simeq0.12/t$. 
	The latter  are defined from a density functional theory calculation performed for \CCOC{} using the Wien2k package~\cite{wien2k,blaha2020}.
	We fit the usual half-filled single band with maximally localized Wannier functions~\cite{marzari1997,souza2001} using the wannier90 code~\cite{mostofi2008,mostofi2014}, and kept the three first hopping terms since the higher-range ones were negligible. 
	The onsite-interaction $U$ is determined from a fit of the measured magnon spectrum. 
	Details are given in Sec.~\ref{sec:results}.
	
	The one-band Hubbard model is then solved using Cluster Dynamical Mean-Field Theory (CDMFT)~\cite{lichtenstein1998,lichtenstein2000,kotliar2001,maier2005,georges1996} with a $2\times2$ plaquette cluster [see Fig.~\ref{intro-pict}(d)].
	We performed antiferromagnetic CDMFT calculations for $x=0$ and $x=0.1$. 
	The calculation was initialized with a polarized and constant self-energy. 
	For the impurity solver we used the hybridization-expansion continuous-time Monte Carlo~\cite{rubtsov2005}  solver~\cite{shinaoka2014} based on the ALPSCore library~\cite{gaenko2017}.
	This solver allows to compute the two-time two-particle Green's function of the $2\times2$ plaquette, from which $S(\bm{Q},\omega)$ can be extracted on the Matsubara axis, and then analytically continued on the real axis using the Maxent package~\cite{levy2017} of the ALPSCore library~\cite{gaenko2017}.
	Due to the small cluster size, only a restricted number of $\bm{Q}$ points are accessible: $(0,0)$, $(\pi,0)$, $(0,\pi)$ and $(\pi,\pi)$.
	
	\begin{figure*}[t]
		\includegraphics[]{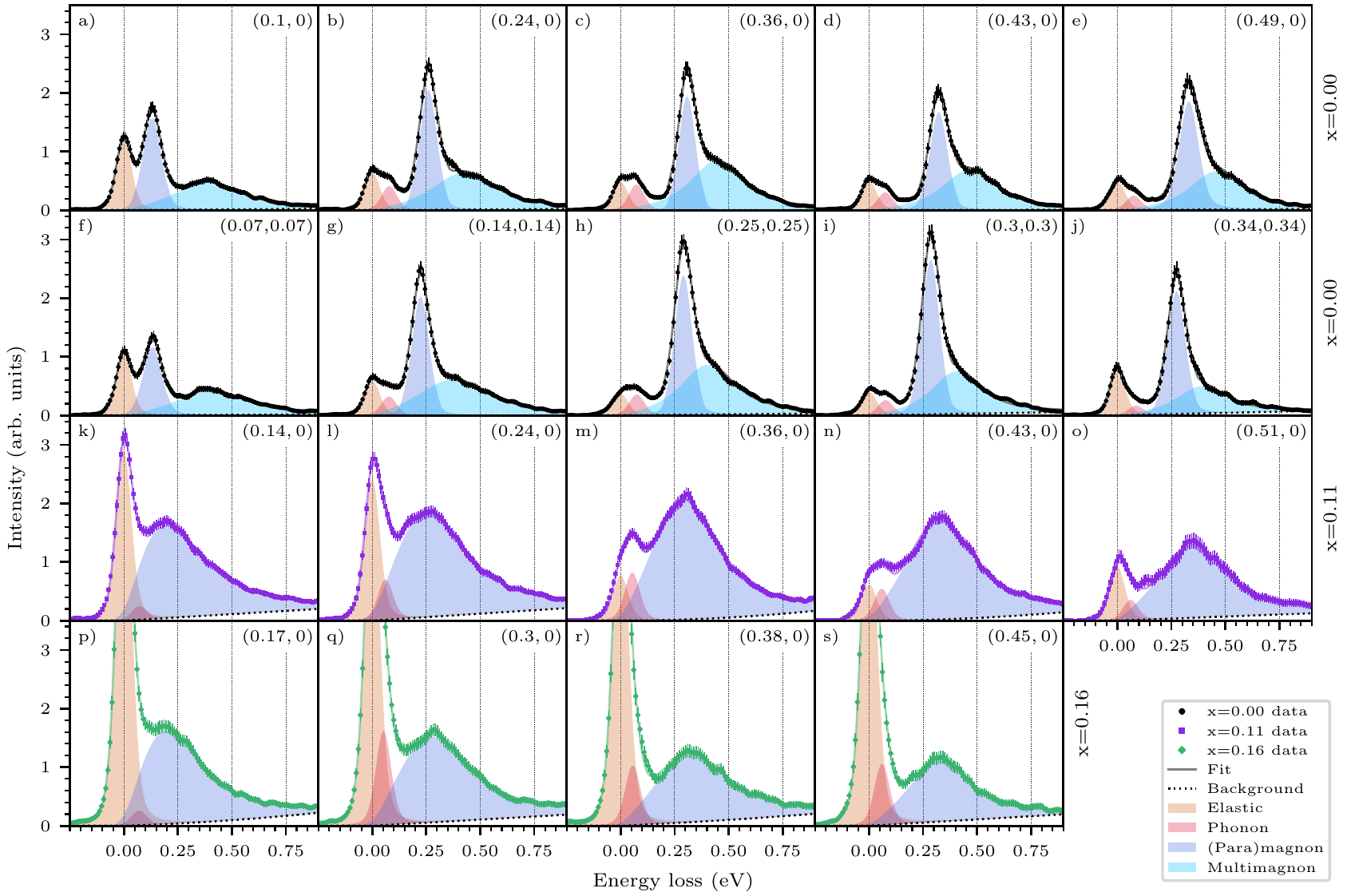}
		\caption{\label{fig_fits} Representative Cu L$_3$-edge RIXS data and fits. 
			Each row corresponds to a specific doping and high-symmetry direction, specifically from top to bottom: $(h, 0)$ for $x=0.00$ [black, (a)-(e)], $(h, h)$ for $x=0.00$ [black, (f)-(j)], $(h, 0)$ for $x=0.11$ [purple, (k)-(o)], and $(h, 0)$ for $x=0.16$ [green, (j)-(n)]. 
			The total fit, background, and individual components correspond respectively to the solid lines, the dotted black lines, and the shaded areas on top of the background (see text for fitting details).
		}
	\end{figure*}
	
	\section{\label{sec:results} Results }
	
	In the full representative spectra shown in Fig.~\ref{RIXS-example} we see all the excitations captured by RIXS in the measured energy window. 
	These include the quasielastic lines and phonons at the lowest energy, magnetic excitations around 0.2-0.4 eV, $dd$ excitations around 1--3~eV, and charge-transfer excitations, clearly visible between 3 to 8~eV, although with some tails possibly down to 2~eV  (left column). 
	The right column of Fig.~\ref{RIXS-example} displays a zoom on the $dd$ excitations, which show a manifold contribution in the undoped, parent compound, which gets broader and less structured upon doping, while also softening. These observations are consistent with previously observed evolution of $dd$ excitations with doping in cuprates \cite{n-mat-dean-lascuo, meyers-lscoPRB,PhysRevB.99.134517}. 
	All spectra have been normalized to the weight of $dd$ excitations in this paper and during analysis, as usually done for RIXS investigations of paramagnons in cuprates \cite{ghiringhelli-prl-rixs-mag,Dean2012,dean-bisco-PRL}. 
    Note that this is appropriate for spectra collected with incident $\sigma$ polarization,  while for $\pi$ polarization this, in principle, does not allow to compare scans performed at different q-points (see e.g. Fig. 3 in \cite{MorettiSala2011}). 
    However, the potential problem does not affect the presented results, since no conclusions are drawn from the analysis of the intensity at different scattering angles. 
	The fitted values of the position and damping parameters remain unchanged without such normalization. 
	
	Representative RIXS spectra in the lower energy region are shown in Fig.~\ref{fig_fits}, measured along ($h$,~0) for all three dopings as well as along ($h$,~$h$) for the undoped sample.
	Each row corresponds to a certain doping/direction and each column to roughly the same momentum transfer magnitude $q_\parallel$. 
	All spectra have been corrected for self-absorption effects using the technique described in Refs.~\cite{PhysRevLett.114.217003, PhysRevLett.123.027001}.
	There are contributions from a quasielastic line and a strong Cu-O bond-stretching phonon in all samples.
	At higher energy, the doped samples show a broad peak corresponding to paramagnon and multiparamagnon excitations, while the undoped sample shows a sharp peak corresponding to a magnon and a broad peak corresponding to multimagnon excitations.
	
	\begin{figure}[h]
		\includegraphics[]{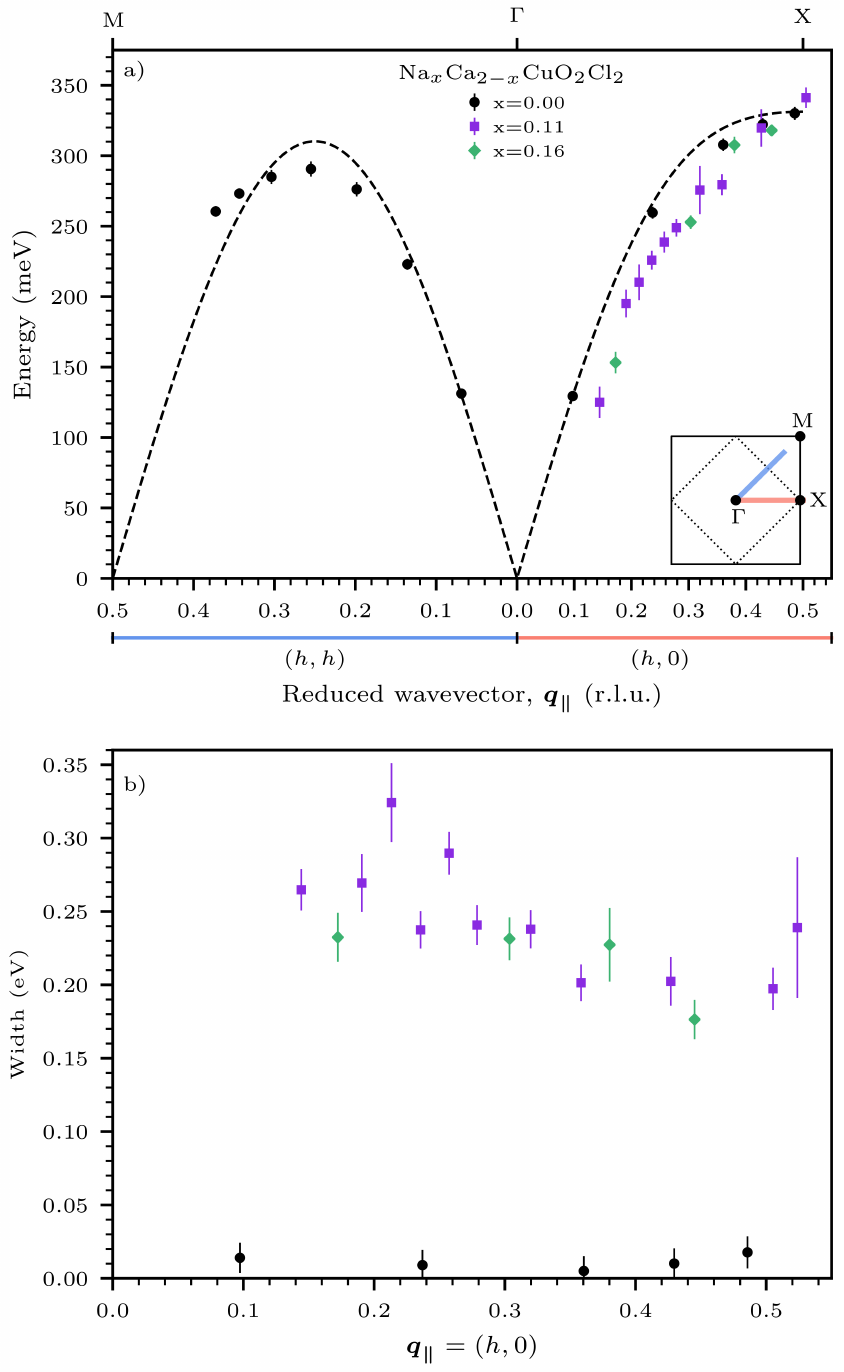}
		\caption{\label{fig_disp_all} (Para)magnon parameters for \NaCCOC{} extracted from Cu L$_3$-edge RIXS data measured at 22~K. (a) Energy dispersion along the two high-symmetry directions as shown by the blue [$(h, h)$] and red [$(h, 0)$] paths in the inset Brillouin zone. The first Brillouin zone boundary is represented by a solid black square, while the magnetic Brillouin zone boundary is represented by an inscribed dotted diamond. The undoped sample (black circle) was measured along both directions, while $x=0.11$ (purple square) and $x=0.16$ (green diamond) were only measured along $(h, 0)$. The fit obtained from the Heisenberg spin-only model is shown as black dashed line (see text for details).  (b) Comparison of extracted HWHM width along $(h, 0)$ for the three different doping levels.}
	\end{figure}
	
	The total fit (solid line), the individual contributions (shaded peaks), as well as a polynomial background (dotted line) are shown in Fig.~\ref{fig_fits}. 
	The quasielastic line and phonon peak are modeled with a Voigt function representing our instrumental resolution. All the magnetic features are modeled with an antisymmetrized Lorentzian with a Bose factor:
	\begin{equation}
		\frac{1}{1-e^{-\epsilon / k_B T}}\frac{A}{\pi}\left[\frac{\gamma}{(\epsilon-\epsilon_0)^2+\gamma^2} - \frac{\gamma}{(\epsilon+\epsilon_0)^2+\gamma^2}\right]
	\end{equation}
	with $\epsilon=\hbar\omega_i-\hbar\omega_f$ the energy loss, $\epsilon_0$ its median value, that we assume as the energy position of the paramagnon,  $\gamma$ the Half Width Half Maximum (HWHM), which is a measure of the excitation damping, and directly connected to the lifetime of the quasiparticle $\tau=\frac{\hbar}{\gamma}$ if a single excitation can be isolated. 
	Finally, $T$ is the measurement temperature, and $A$ the intensity. The antisymmetrized Lorentzian is numerically convolved with a Voigt resolution function. 
    We confirmed that using a Damped-Harmonic-Oscillator (DHO) model gives the same results within errorbars. 
    The DHO model is sometime considered to better account for the physical lineshape \cite{PhysRevB.93.214513,PhysRevB.98.144507}, in particular in the overdamped regime. 
    Yet, we note first that this is still an empirical approach with no fundamental justification. 
    Moreover, in the present case, for the undoped sample with long range antiferromagnetic order the single magnon contribution was well separated from the multi-magnon ones, and with a small damping.
    In the doped case, it was not possible to separate the 2 contributions, and fitting with a DHO would have implied a strong assumption on the single magnon line-shape. 
    Our simplified approach based on inspecting the position of the Lorentzian peak allows a comparison with a minimum of assumptions.

	\begin{figure}[h]
		\includegraphics[]{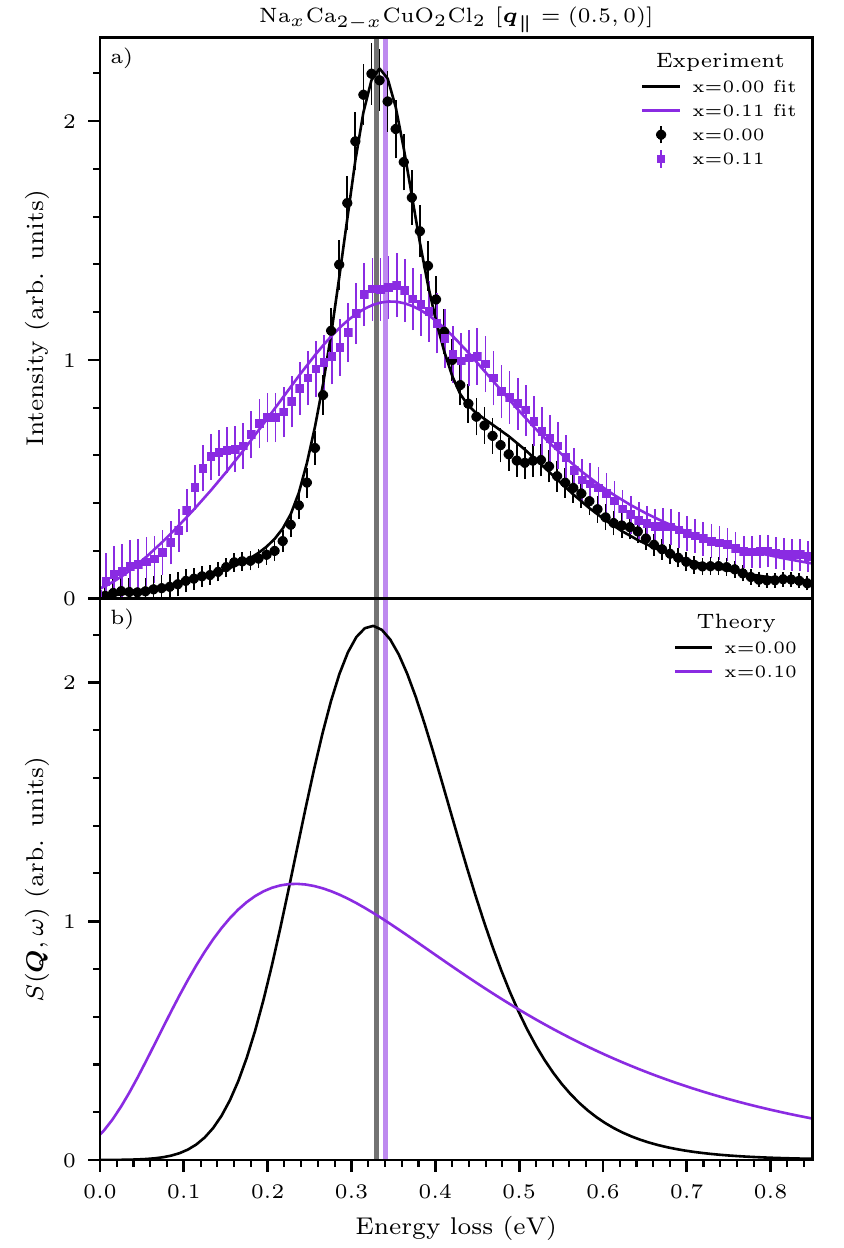}
		\caption{\label{fig_theory} Comparison between experiment and theory at $\boldsymbol{q}_{\boldsymbol{\parallel}}=(0.5, 0)$ for \NaCCOC{}. (a) Cu L$_3$-edge RIXS data for $x=0.00$ at $\boldsymbol{q}_{\boldsymbol{\parallel}}=(0.49, 0)$  (black circle) and $x=0.11$ at $\boldsymbol{q}_{\boldsymbol{\parallel}}=(0.51, 0)$ (purple square). The fits are shown as solid lines in black ($x=0.00$) and purple ($x=0.11$). The background, phonon, and elastic contributions have been subtracted from both the data and fits to compare with theory. The energy of the fitted magnon (paramagnon) at $x=0.00$ ($x=0.11$) is shown as a vertical black (purple) line. (b) CDMFT calculations at $x=0.00$ (black) and $x=0.1$ (purple). The calculated spectra have been multiplied by a Bose factor with experimental temperature (22~K) and convolved with the experimental resolution function (83~meV FWHM).}
	\end{figure}

	The parameters from the fits of the (para)magnons are summarized in Fig.~\ref{fig_disp_all} for the energy position (a) and the HWHM width (b).
	The maximum energy for the undoped, antiferromagnetic \CCOC{} is found to be \Emax=$330 \pm 4$ meV at the zone boundary $\boldsymbol{q}_{\boldsymbol{\parallel}}=(0.5, 0)$, while the $x = 0.11$ doped case yields \Emax=$341 \pm 10$ meV, and in general around   $\boldsymbol{q}_{\boldsymbol{\parallel}}=(0.4, 0)$ to $(0.5, 0)$ the energies of both the doped and the undoped are within error bars, while at lower wavevector along the $\Gamma -M$ line, for $\boldsymbol{q}_{\boldsymbol{\parallel}}\leq(0.35, 0)$, we observe a distinct softening of the paramagnons in the two doped samples compared to the magnons in the undoped one.    

	The on-site Hubbard interaction $U$ can be determined from the measured magnon spectrum.
	Indeed, based on the observation that \CCOC{} displays magnetic order~\cite{ohishi2005}, we fit the magnon dispersion in undoped samples using the dispersion relation of a Heisenberg spin-only model [see Fig.~\ref{intro-pict}(b)]:
	\begin{equation}
		\label{eq:Hspin}
		\begin{split}
			H_{\mathrm{spin}} =  &\mathcal{J}\sum_{\langle ij\rangle}\mathbf{S}_{i}\mathbf{S}_j + \mathcal{J}\p\sum_{\langle\langle ij\rangle\rangle}\mathbf{S}_{i}\mathbf{S}_j+ \mathcal{J}\pp\sum_{\langle\langle\langle ij\rangle\rangle\rangle}\mathbf{S}_{i}\mathbf{S}_j \\
			& + \mathcal{J}_c\sum_{\langle ijkl\rangle}\bigg[\left(\mathbf{S}_{i}\mathbf{S}_j\right)\left(\mathbf{S}_{k}\mathbf{S}_l\right)+\left(\mathbf{S}_{i}\mathbf{S}_l\right)\left(\mathbf{S}_{k}\mathbf{S}_j\right)\\
			& -\left(\mathbf{S}_{i}\mathbf{S}_k\right)\left(\mathbf{S}_{j}\mathbf{S}_l\right)\bigg]
		\end{split}
	\end{equation}
	where $\langle ijkl\rangle$ refers to a square plaquette of four neighboring sites $i,j,k,l$ [see Fig.~\ref{intro-pict}(b)], and the exchange couplings are linked to the Hubbard terms of Eq. \ref{eq:Hubbard} as follows: $\mathcal{J} = \frac{4t^2}{U} - \frac{24t^4}{U^3}$ which is the NN main Heisenberg superexchange parameter, $\mathcal{J}\p=\frac{{4t\p}^2}{U}+\frac{4t^4}{U^3}$ the next-NN one, $\mathcal{J}\pp=\frac{{4t\pp}^2}{U}+\frac{4t^4}{U^3}$ the next-next-NN and $\mathcal{J}_{c}=\frac{80t^4}{U^3}$ the cyclic exchange term  ~\cite{coldea-prl-lco,delannoy2009} (see Fig. \ref{intro-pict}(b)).
	Since the hopping parameters are fixed to the values obtained \emph{ab initio}, $U$ is the only free parameter.
	The resulting fit is shown in Fig.~\ref{fig_disp_all} as a dashed line. 
	The cyclic exchange term is essential to account for the energy difference between $(0.5,0)$ and $(0.25,0.25)$. 
	We find $U\simeq4.34\si{eV}\simeq10.2t$, in good agreement with the usual one-band model parameterizations for cuprates found from \emph{ab initio} calculations~\cite{hirayama2018,moree2022} (see Table~\ref{tab:tableHubbard}).
	Note that we neglected the contributions of additional ring exchange terms which are of the type $t^{2}t\p t\pp/U^{3}$.
	The correction from these higher order terms would be $\sim$1 meV, much smaller than the experimental resolution, while the leading term for the cyclic exchange $\mathcal{J}_c=80t^{4}/U^{3}$ is about 30 meV.

	In Fig. \ref{fig_theory}(a), we focus on the data near the zone boundary $\bm{q_\parallel} \simeq (0.5,0)$ for the undoped (black circles) and x $=$ 0.11 doped (purple circles) samples, where the background, quasielastic, and phonon contributions determined from our fit have been removed in order to focus only on the paramagnon contribution. 
	In Fig. \ref{fig_theory}(b), we show the calculated dynamical spin structure factor using cluster DMFT, as described above (Sec.  \ref{Met:CDMFT}), at similar wavevector and doping. 
	A vertical bar shows the fitted energy position for the undoped (black) and doped (purple) experimental data. 
	
	\section{\label{sec:discu} Discussion}
	
	The dispersion of the magnetic excitations, as shown in Fig. \ref{fig_disp_all}, panel a), gives important information about a system's electronic and magnetic states. 
	Furthermore, for compounds that magnetically order, the magnon dispersion provides a direct access to the superexchange since it can be reliably modeled with a simple Heisenberg Hamiltonian. 
	This has been done in all cuprate parent compounds, where the dispersion along the main in-plane symmetry directions can be modeled taking into account the superexchange parameters up to the next-next-NN as well a cyclic one within the CuO square plaquette\cite{coldea-prl-lco}. 
	In a previous work~\cite{ccoc-sw-blair1} Lebert \textit{et al.} already attempted such modeling for the parent, antiferromagnetic compound \CCOC{}. 
	The dispersion along the $\Gamma-X$ (antinodal) direction, gave the maximum energy and is determined mainly by the leading NN exchange term $\mathcal{J}$, while the difference between the antinodal and nodal ($\Gamma$-M) direction is determined by the other terms $\mathcal{J}\p$,  $\mathcal{J}\pp$ and $\mathcal{J}_c $. 
	This approach works well to determine $\mathcal{J}$, but has the limitation that several parameter sets can approximately match the experimental dispersion (see, \textit{e.g.} Fig. 3 in Dalla Piazza \textit{et al.} Ref. \onlinecite{dalla_piazza_rapidB}). 
	%
	%
	
	%
	\begin{table}[t]
		\caption{\label{tab:tableHubbard}%
			Comparison of hopping and Hubbard on-site interaction terms between the present work and previous ones on similar compounds. The estimates in the first part of the tables are obtained from the experimental data while those in the second part are entirely determined from \emph{ab initio} calculations.
		}
		\begin{ruledtabular}
			\begin{tabular}{lcccr}
				&
				$t (\si{meV})$ &
				$\left|t\p/t\right|$ &
				$\left|t\pp/t\right|$ &
				$U/t$\\
				\colrule\\
				\multicolumn{5}{c}{\emph{Ab initio} calculations and experiment combined.} \\ 
				\multicolumn{5}{c}{} \\
				CCOC (This work) & 425 & 0.18 & 0.12 & 10.2\\ 
				\colrule\\
				\multicolumn{5}{c}{Fitted on experimental data.} \\ 
				\multicolumn{5}{c}{} \\
				CCOC \cite{ccoc-sw-blair1} & 295 & -- & -- & 7.46\\ 
				SCOC \cite{dalla_piazza_rapidB} & 480(10) & 0.42 & 0.16 & 7.29\\   
				SCOC \cite{PhysRevLett.80.4245}\footnote{These hopping parameters were derived from a $t-J$ model.} & 350 & 0.34 & 0.23 & --\\    
				LCO \cite{dalla_piazza_rapidB} & 492(7) & 0.42 & 0.09 & 7.11\\
				BSYCO \cite{dalla_piazza_rapidB} & 470  & 0.44 & 0.17 & 7.44 \\   
				BSCCO \cite{dean-bisco-PRL}$^{\textcolor{blue}{\text{a}}}$& 144.0  & 0.3  & 0.2 & -- \\   
				BSCO \cite{peng2017}\footnote{Note that the exchange contributions shown here are the ones obtained from a single band Hubbard model with NN hopping $t$. For an effective Heisenberg model up to 6-NN, see Ref. \onlinecite{peng2017}.} & 270  & --  & -- & 6.2 \\      
				CCO \cite{peng2017}$^{\textcolor{blue}{\text{b}}}$  & 297  & --  & -- & 4.9 \\  
				\colrule \\
				\multicolumn{5}{c}{\emph{Ab intio} calculations only.} \\ 
				\multicolumn{5}{c}{} \\
				LCO \cite{hirayama2018} & 482 & 0.15 & 0.21 & 10.36  \\
				HBCO \cite{hirayama2018} & 461 & 0.26 & 0.16 & 9.49  \\
				HBCO \cite{moree2022} & 494 & 0.23 & 0.11 & 8.16  \\
				CCO \cite{moree2022} & 521 & 0.23 & 0.06 & 8.6  \\
				BSCO \cite{moree2022}\footnote{These estimates were obtained at an effective $x=0.2$ doping.} & 527 & 0.27 & 0.08 & 8.34 \\
				BSCCO \cite{moree2022}$^{\textcolor{blue}{\text{c}}}$ & 451 & 0.29 & 0.11 & 9.37
			\end{tabular}
		\end{ruledtabular}
	\end{table}
	In the present work, we measured at a higher resolution which allows us to clearly separate the single magnon from the multimagnon contribution in the undoped, anti-ferromagnetic compound, as shown in Fig. \ref{fig_fits}, panels a-j), and \ref{fig_theory}, panel a). 
	The line-shape is similar to the one expected in anti-ferromagnetic cuprates with incident light in $\pi$ polarization, and closely resembles the one in \SCOC{}, where a polarimetry analysis confirms our identification in terms of single and multi-magnon contributions\cite{PhysRevB.103.L140409}.     
	We also compare these results with two dopings, in the superconducting region, for x $=$ 0.1 and 0.16, although only along the antinodal direction $\Gamma-X$. 
	We find that the doping does not affect the dispersion greatly, although in Fig.~\ref{fig_disp_all}(a) a small softening with doping is observed for $h \leq 0.3$, while recovering to the same energy at the zone boundary $X=(0.5,0)$. 
	This has already been observed in hole-doped cuprates, and interpreted in terms of a tendency toward ferromagnetic correlation of the spins upon doping\cite{Jia2014}. 
	We note, however, that in the doped case a separation of the paramagnon from the multiparamagnon contributions is impossible due to broadening, as shown in Figure \ref{fig_fits} and \ref{fig_theory}.
	
	To interpret the observed dispersion we used an approach, in which the Hubbard $U$ is determined from a fit of the experimental magnon dispersion with fixed hopping parameters obtained \emph{ab initio} (i.e. $U$ is the only free parameter).
	The results of such approach are compared in Table~\ref{tab:tableHubbard} to fit on previous RIXS experimental works on undoped 
	\CCOC{} (CCOC) from Ref. \onlinecite{ccoc-sw-blair1}, and on single layers 
	\SCOC{} (SCOC), \LCO{} (LCO), from Ref.~\onlinecite{dalla_piazza_rapidB} and
	\BSCO{} (BSCO) from Ref. \onlinecite{peng2017}, 
	as well as double layer \BYSCO ~(BSYCO) also from Ref.~\onlinecite{dalla_piazza_rapidB} and \BiTTOT{} (BSCCO) from Ref. \onlinecite{dean-bisco-PRL},
	and finally infinite layer \CCO{} (CCO) again from Ref. \onlinecite{peng2017}. 
	We also add a fit on ARPES data on SCOC from Ref. \onlinecite{PhysRevLett.80.4245}.
	In Ref.~\onlinecite{dalla_piazza_rapidB} their approach was to simultaneously fit several compounds to obtain a common set of Hubbard parameters, but, opposite to our present approach, fixing the value of U while fitting simultaneously the three hopping parameters. Note also that in Ref.~\onlinecite{ccoc-sw-blair1}, the 4 Heisenberg parameters are obtained by keeping only the NN hopping term $t$. We also compare to fully \emph{ab initio}-derived single band models for LCO, \HBCO{} (HBCO),  \CCO{} (CCO), \BSCO{} (BSCO) and \BSCCO{} (BSCCO)\cite{hirayama2018,moree2022}.

	\begin{table}[t]
		\caption{\label{tab:Heisenberg}%
			Comparison of Heisenberg interaction terms between the present work and previous ones on similar compounds.  
		}
		\begin{ruledtabular}
			\begin{tabular}{lcccr}
				&
				$\mathcal{J} (\si{meV})$ &
				$\mathcal{J}\p (\si{meV}$) &
				$\mathcal{J}\pp (\si{meV}$) &
				$\mathcal{J}_c (\si{meV}$)\\
				\colrule\\
				CCOC (This work) & 157.1 & 6.9 & 3.9 & 32.0\\ 
				\colrule\\
				CCOC \cite{ccoc-sw-blair1}\footnote{Note that in Ref.~\onlinecite{ccoc-sw-blair1} all exchange contributions are obtained from a single t value.} & 141.2 & 2.8 & 2.8 & 56.9\\ 
				LCO \cite{dalla_piazza_rapidB} \footnote{ data from \cite{coldea-prl-lco}} & 140 & -- & -- & --\\   
				SCOC \cite{dalla_piazza_rapidB}\footnote{data from \cite{guarise-rixs-mag-prl}} & 120 & -- & -- & --\\   
				SCOC \cite{PhysRevLett.80.4245} & 140 & -- & -- & --\\   
				BSYCO \cite{dalla_piazza_rapidB} &  150 & -- & -- & --\\   
				BSCCO \cite{dean-bisco-PRL} & 120  & --  & -- & -- \\     
				BSCO \cite{peng2017}\footnote{Note that the exchange contributions shown here are the ones obtained from a single band Hubbard model with NN hopping $t$. For an effective Heisenberg model up to 6-NN, see Ref. \onlinecite{peng2017}.} & 148  & --  & -- & 41 \\      
				CCO \cite{peng2017}$^{\textcolor{blue}{\text{d}}}$  & 192  & --  & -- & 206 \\  
			\end{tabular}
		\end{ruledtabular}
	\end{table}

	It is encouraging to observe that our hybrid method for determining the Hubbard $U$ is in line with either fully \emph{ab initio} calculations of Ref.~\cite{hirayama2018,moree2022}, and the fitting procedures of experimental data performed in Ref.~\cite{dalla_piazza_rapidB,PhysRevLett.80.4245,dean-bisco-PRL}.
	For the purpose of our fitting procedure, we assume
	purely local Coulomb interactions, that is, we only use the local on-site U in our fit. 
	For cuprates, the longer-range interaction terms are rather small, which is why our values of U are still in good agreement, even if slightly overestimated, with those obtained purely \textit{ab initio} \cite{hirayama2018,moree2022}. Indeed, the constrained RPA
	technique \cite{Aryasetiawan2004} used in the ab initio works
	a priori includes the full (partially) screened Coulomb interactions,
	including its local and non-local components.
	We note that the ratio $\left|t/t\p\right|$ found by Dalla Piazza \emph{et al.} are systematically larger than the \emph{ab initio} parameters, and than the ratio we found for CCOC. 
	Since the estimations reported in Table~\ref{tab:tableHubbard} come from a different fitting procedure, which can lead to variations in the value of the hopping parameters, we also compare the Heisenberg terms themselves in Table~\ref{tab:Heisenberg}.
	The value of the Heisenberg term should be dictated by the experimental measurements, and therefore less sensitive to the methodological differences. 
	One can observe that, the present values of $\mathcal{J}$ are not too far from the previous fit on CCOC~\cite{ccoc-sw-blair1}, which in turn was matching previous results on others cuprates. 
	As expected, this is the case despite the fact that the Hubbard $U$ and $t$ on the contrary are quite far, the difference being larger than for the same Hubbard values $U$ and $t$ on Ref. \onlinecite{dalla_piazza_rapidB}.
	The large difference in $U$ is mainly a direct consequence of the difference in the hopping $t$: $\mathcal{J}$ should roughly be the same to match the experimental data, so that if $t$ decreases then $U/t$ has to decrease to obtain a similar exchange value since $\mathcal{J}\propto t\frac{t}{U}$. 
	Hence our approach is an accurate and computationally cheap way of extracting reliable parameters for low-energy effective Hamiltonians. The non-uniqueness of the parameter sets for a same $\mathcal{J}$ is overcome by fixing the hopping parameters according to an \emph{ab initio} calculation.
	It could be used in other cuprates in order to attempt a unified one-band Hubbard model for magnetic and electronic spectra. 
	The energy dispersion along the zone boundary in the undoped \CCOC{}, as estimated from the difference between  the energy at $\boldsymbol{q}_{\boldsymbol{\parallel}}=(0.5, 0)$ and $(0.25, 0.25)$ is 40$\pm$8 meV, which is within error bars identical to the one of undoped \LCO. 
	Indeed, we found a rather close value for the cyclic Heisenberg term $\mathcal{J}_c $ as this parameter determines the energy dispersion along the zone boundary \cite{Headings2010,peng2017}. 
	Surprisingly, this is generally considered to vary with the apical oxygen coordination, which is, however, in our case replaced by a chlorine ion, so that one would expect rather different zone boundary dispersion  \cite{peng2017}, which is clearly not the case here.

	Finally, several works have suggested that the superexchange parameter $\mathcal{J}$ could give a reasonable estimate of the order of magnitude of the superconducting transition temperature $T_\mathrm{c}$ \cite{tacon-paramag, ivashko2019, wang2022}. 
	In particular Wang \textit{et al.}  show in Fig.~4(d) of their work a possible linear relation between $T_{\text{c,max}}$ and $\mathcal{J}$ for various compounds~\cite{wang2022}.
	In line with other studies in different cuprate families, \cite{PhysRevB.90.220506,PhysRevB.92.104507} we find that oxychloride cuprates, marked as CCOC in Fig.~4(d) of Ref. \onlinecite{wang2022}, strongly deviate from this relation.
	The maximum $T_\mathrm{c}$ reported for CCOC is $\sim 40$~K \cite{Yamada2005}, which differs from the value displayed in Ref. \onlinecite{wang2022}.
	We find $\mathcal{J}\approx 157.1$ meV, such that CCOC would match the values of \LaTOF~ in the same plot of Ref. \onlinecite{wang2022}.
	Hence CCOC lies far away from the linear relation proposed. 
	Our analysis also suggests that the conclusions of Ref. \onlinecite{ivashko2019} regarding the link between $T_\mathrm{c}$ and $\mathcal{J}$ may not be naively extrapolated across different cuprate compounds.

	A second important parameter that we could obtain, is the evolution of the paramagnon excitation broadening with doping, as shown in Fig. \ref{fig_disp_all}(b). 
	This is a key parameter to understand the evolution of the spin fluctuations upon adding free carriers. 
	In self-consistent renormalization spin theory \cite{Moriya_2003}, the frequency spread of the spin fluctuation is even directly linked to the superconducting $T_\mathrm{c}$. 
	Contrary to the energy position, the width changes strongly upon doping as seen in Fig.~\ref{fig_disp_all}(b). 
	To understand better the evolution of the paramagnon broadening, we model the full dynamical spin structure factor using cluster DMFT on a 4-site plaquette as described in Sec.~\ref{Met:CDMFT}, for the undoped and x $=$ 0.1 doping. 
	This effectively limited us to the zone boundary, that we could measure only at the $X$ point (0.5, 0), as shown in Fig.~\ref{fig_theory}. 
	We find a very good match with the energy position in the undoped, antiferromagnetic case, and a reasonably good match with the spectral shape. 
	
	This is true only for calculations in the antiferromagnetic state, since the paramagnetic one gave a sizable softening relative to experiment even for the undoped case. 
	For the doped case, we find a very good match for the evolution of the area, that is controlled by the increased width. 
	Indeed, the decrease in intensity is the same as in the experiment to a very good approximation. 
	Incidentally, we note that the FWHM of the spin excitation on the doped samples is $395 \pm 9$ meV at $\boldsymbol{q}_{\boldsymbol{\parallel}}=(0.51, 0)$ experimentally, while the FWHM is $614 \pm 5$ meV for the model at $\boldsymbol{q}_{\boldsymbol{\parallel}}=(0.5, 0)$.
	This will give a characteristic temperature indicating the energy spread of the wave vector-dependent part of the spin fluctuations $T_0=4600 \pm 100$ K. 
	The sample shows a $T_\mathrm{c}\approx$ 14 K, which lies quite far below from the logarithmic relation $T_\mathrm{c}(T_0)$ shown in Fig. 6 of Ref. \onlinecite{Moriya_2003}, compared to other cuprates superconductors.
	However, we observe a softening of the energy position of the maximum in the model, while the experiment shows about the same position. 
	This softening is not linked to the analytic continuation since the same parameters were used for all dopings.
	We checked that modifications of the default model used in the analytic continuation procedure does not lead to a significant change in the peak position. 
	The softening is interpreted as being a consequence of the melting of the \emph{in-plane} antiferromagnetic correlations upon doping, that does not happen in the real material. 
	Indeed, a similar softening is observed (not shown) in the undoped case when increasing the temperature such that the antiferromagnetic order melts, while it is well known that the high energy part of the dispersion is stable over a larger temperature range \cite{coldea-prl-lco}. 
	While the self-energy is almost purely static in the ordered phase, a strong dynamic component develops upon doping or increasing temperature. 
	This points to the conclusion that CDMFT calculations of the one-band Hubbard model on small cluster requires a quasi-static order to reproduce faithfully the RIXS measurements. 
	
	One possible origin of this discrepancy could be that it is a limitation of the one-band Hubbard model. 
	However, this option can be discarded since determinant quantum Monte Carlo calculations showed that both the one-band\cite{Jia2014} and the three-band\cite{chaix2018} models capture the absence of softening at $X=(0.5,0)$.
	These calculations were performed on a substantially larger system ($8\times8$) as compared to the $2\times2$ plaquette we use in this work, which suggests that the softening at $X=(0.5,0)$ is most probably an effect of the small cluster size.
	While upon electron-doping it has been identified that the three-site hopping term in the $t-\mathcal{J}$ model can account for the absence of softening, the same conclusion can not be drawn for hole-doping\cite{Jia2014}.
	Our calculations show that although magnetic correlations are strongly suppressed when adding holes, it is still necessary to account for longer-range magnetic correlations beyond those that are captured by the $2\times2$ plaquette cluster.
	
	\section{Conclusions}
	
	Using Resonant Inelastic X-ray Scattering, we measure the paramagnon dispersion and damping of undoped, antiferromagnetic \CCOC{} as well as doped, superconducting \NaCCOC{}. 
	In combination with these measurements we perform calculations of the dynamical spin structure factor of the one-band Hubbard model using cluster dynamical mean-field theory. 

	A first major result of this work is that we could extract the \emph{on-site} Coulomb repulsion parameter $U\simeq4.34$ $\si{eV}$ ($U/t$=10.2) for a Hubbard Hamiltonian of \CCOC, as the single fitting parameter of the paramagnon dispersion. 
	We confirmed that the obtained parameters for the effective low-energy model is in agreement with usual parameterizations for cuprates~\cite{hirayama2018,moree2022}. 
	This is of general interest since it is a computationally cheap method to obtain at least the order of magnitude of the Hubbard $U$ given the experimental magnon dispersion.
	The latter is characterized by a maximum energy for the undoped, antiferromagnetic \CCOC{} of \Emax=$330 \pm 4$~meV at the zone boundary $X=(0.5, 0)$, and of \Emax=$341 \pm 10$~meV for the $x = 0.11$ doped case.  
	From the fit of the overall dispersion with an Heisenberg model, we find a superexchange $\mathcal{J}$ of 157.1 meV, relatively high in comparison with the low $T_\mathrm{c}\sim14-16~\si{K}$.
	In contrast to the relatively low $T_\mathrm{c}$ of Na-CCOC, our measurements show that the superconducting critical temperature may not be straightforwardly related to $\mathcal{J}$.
	The analysis of the magnon width at the zone boundary $X=(0.5,0)$ also sheds uncertainty on the heuristic correlation proposed by Moriya \emph{et al.}~\cite{Moriya_2003}.
	These observations are confirmed by our cluster dynamical mean-field theory calculations on a $2\times2$ plaquette. 
	Both the undoped magnon spectrum  and the increase in width at the zone boundary is well captured by the theory.
	Moreover, our calculations clearly show that upon doping only short-range spin fluctations are not sufficient to capture the spin fluctuation spectrum, and hence provide precious information for further studies. 
	
	\begin{acknowledgments}
		The authors are grateful to Beno\^it Baptiste and Lise-Marie Chamoreau for their assistance in crystal orientation and acknowledge the use of the  x-ray diffractometer instrument at the ''Plateforme Diffraction'', IPCM, and IMPMC Paris.
		M.d'A. is very grateful to Marie-Aude M\'easson, Giacomo Ghiringhelli and Chafic Fawaz for critical reading of the manuscript. 
		B.W.L.\ acknowledges financial support from the French state funds managed by the ANR within the ``Investissements d'Avenir'' programme under reference  ANR-11-IDEX-0004-02, and within the framework of the Cluster of Excellence MATISSE led by Sorbonne Universit\'{e} and from the LLB/SOLEIL PhD fellowship program. Work at Brookhaven National Laboratory was supported by the U.S. Department of Energy (DOE), Office of Science, Office of Basic Energy Sciences.
		B.B.-L.\ acknowledges funding through the Institut Polytechnique de Paris.
		S.H. was supported by Academy of Finland (project 295696).
		We acknowledge the MPBT platform (Sorbonne University) for the use of its SQUID magnetometer.
		We acknowledge syncrotron beam-time and experiment financial support to ESRF trough experiment HC-2702.
		We acknowledge supercomputing time at the French Grand Equipement National de Calcul Intensif IDRIS-GENCI Orsay (Project No. A0130901393) and we thank the CPHT computer support team. 
	\end{acknowledgments}


\begin{thebibliography}{76}
\expandafter\ifx\csname natexlab\endcsname\relax\def\natexlab#1{#1}\fi
\expandafter\ifx\csname bibnamefont\endcsname\relax
  \def\bibnamefont#1{#1}\fi
\expandafter\ifx\csname bibfnamefont\endcsname\relax
  \def\bibfnamefont#1{#1}\fi
\expandafter\ifx\csname citenamefont\endcsname\relax
  \def\citenamefont#1{#1}\fi
\expandafter\ifx\csname url\endcsname\relax
  \def\url#1{\texttt{#1}}\fi
\expandafter\ifx\csname urlprefix\endcsname\relax\def\urlprefix{URL }\fi
\providecommand{\bibinfo}[2]{#2}
\providecommand{\eprint}[2][]{\url{#2}}

\bibitem[{\citenamefont{Bednorz and M{\"u}ller}(1986)}]{Bednorz1986}
\bibinfo{author}{\bibfnamefont{J.~G.} \bibnamefont{Bednorz}} \bibnamefont{and}
  \bibinfo{author}{\bibfnamefont{K.~A.} \bibnamefont{M{\"u}ller}},
  \bibinfo{journal}{Zeitschrift für Physik B Condensed Matter}
  \textbf{\bibinfo{volume}{64}}, \bibinfo{pages}{189} (\bibinfo{year}{1986}),
  ISSN \bibinfo{issn}{1431-584X},
  \urlprefix\url{https://doi.org/10.1007/BF01303701}.

\bibitem[{\citenamefont{Scalapino}(1995)}]{Scalapino1995329}
\bibinfo{author}{\bibfnamefont{D.}~\bibnamefont{Scalapino}},
  \bibinfo{journal}{Physics Reports} \textbf{\bibinfo{volume}{250}},
  \bibinfo{pages}{329 } (\bibinfo{year}{1995}), ISSN \bibinfo{issn}{0370-1573},
  \urlprefix\url{http://www.sciencedirect.com/science/article/pii/037015739400086I}.

\bibitem[{\citenamefont{Moriya and Ueda}(2003)}]{Moriya_2003}
\bibinfo{author}{\bibfnamefont{T.}~\bibnamefont{Moriya}} \bibnamefont{and}
  \bibinfo{author}{\bibfnamefont{K.}~\bibnamefont{Ueda}},
  \bibinfo{journal}{Reports on Progress in Physics}
  \textbf{\bibinfo{volume}{66}}, \bibinfo{pages}{1299} (\bibinfo{year}{2003}),
  \urlprefix\url{https://doi.org/10.1088/0034-4885/66/8/202}.

\bibitem[{\citenamefont{Kirtley et~al.}(2006)\citenamefont{Kirtley, Tsuei,
  Ariando, Verwijs, Harkema, and Hilgenkamp}}]{Kirtley-NatPhys2006}
\bibinfo{author}{\bibfnamefont{J.~R.} \bibnamefont{Kirtley}},
  \bibinfo{author}{\bibfnamefont{C.~C.} \bibnamefont{Tsuei}},
  \bibinfo{author}{\bibnamefont{Ariando}},
  \bibinfo{author}{\bibfnamefont{C.~J.~M.} \bibnamefont{Verwijs}},
  \bibinfo{author}{\bibfnamefont{S.}~\bibnamefont{Harkema}}, \bibnamefont{and}
  \bibinfo{author}{\bibfnamefont{H.}~\bibnamefont{Hilgenkamp}},
  \bibinfo{journal}{Nature Physics} \textbf{\bibinfo{volume}{2}},
  \bibinfo{pages}{190} (\bibinfo{year}{2006}), ISSN \bibinfo{issn}{1745-2481},
  \urlprefix\url{https://doi.org/10.1038/nphys215}.

\bibitem[{\citenamefont{Le~Tacon et~al.}(2011)\citenamefont{Le~Tacon,
  Ghiringhelli, Chaloupka, Moretti~Sala, Hinkov, Haverkort, Minola, Bakr, Zhou,
  Blanco-Canosa et~al.}}]{tacon-paramag}
\bibinfo{author}{\bibfnamefont{M.}~\bibnamefont{Le~Tacon}},
  \bibinfo{author}{\bibfnamefont{G.}~\bibnamefont{Ghiringhelli}},
  \bibinfo{author}{\bibfnamefont{J.}~\bibnamefont{Chaloupka}},
  \bibinfo{author}{\bibfnamefont{M.}~\bibnamefont{Moretti~Sala}},
  \bibinfo{author}{\bibfnamefont{V.}~\bibnamefont{Hinkov}},
  \bibinfo{author}{\bibfnamefont{M.~W.} \bibnamefont{Haverkort}},
  \bibinfo{author}{\bibfnamefont{M.}~\bibnamefont{Minola}},
  \bibinfo{author}{\bibfnamefont{M.}~\bibnamefont{Bakr}},
  \bibinfo{author}{\bibfnamefont{K.~J.} \bibnamefont{Zhou}},
  \bibinfo{author}{\bibfnamefont{S.}~\bibnamefont{Blanco-Canosa}},
  \bibnamefont{et~al.}, \bibinfo{journal}{Nat Phys}
  \textbf{\bibinfo{volume}{7}}, \bibinfo{pages}{725} (\bibinfo{year}{2011}),
  \urlprefix\url{http://dx.doi.org/10.1038/nphys2041}.

\bibitem[{\citenamefont{Wang et~al.}(2022)\citenamefont{Wang, He, Yang,
  Garcia-Fernandez, Nag, Zhou, Minola, Tacon, Keimer, Peng et~al.}}]{wang2022}
\bibinfo{author}{\bibfnamefont{L.}~\bibnamefont{Wang}},
  \bibinfo{author}{\bibfnamefont{G.}~\bibnamefont{He}},
  \bibinfo{author}{\bibfnamefont{Z.}~\bibnamefont{Yang}},
  \bibinfo{author}{\bibfnamefont{M.}~\bibnamefont{Garcia-Fernandez}},
  \bibinfo{author}{\bibfnamefont{A.}~\bibnamefont{Nag}},
  \bibinfo{author}{\bibfnamefont{K.}~\bibnamefont{Zhou}},
  \bibinfo{author}{\bibfnamefont{M.}~\bibnamefont{Minola}},
  \bibinfo{author}{\bibfnamefont{M.~L.} \bibnamefont{Tacon}},
  \bibinfo{author}{\bibfnamefont{B.}~\bibnamefont{Keimer}},
  \bibinfo{author}{\bibfnamefont{Y.}~\bibnamefont{Peng}}, \bibnamefont{et~al.},
  \bibinfo{journal}{Nature Communications} \textbf{\bibinfo{volume}{13}},
  \bibinfo{pages}{3163} (\bibinfo{year}{2022}), ISSN \bibinfo{issn}{2041-1723},
  \urlprefix\url{https://doi.org/10.1038/s41467-022-30918-z}.

\bibitem[{\citenamefont{Kowalski et~al.}(2021)\citenamefont{Kowalski, Dash,
  Sémon, Sénéchal, and Tremblay}}]{kowalski2021}
\bibinfo{author}{\bibfnamefont{N.}~\bibnamefont{Kowalski}},
  \bibinfo{author}{\bibfnamefont{S.~S.} \bibnamefont{Dash}},
  \bibinfo{author}{\bibfnamefont{P.}~\bibnamefont{Sémon}},
  \bibinfo{author}{\bibfnamefont{D.}~\bibnamefont{Sénéchal}},
  \bibnamefont{and} \bibinfo{author}{\bibfnamefont{A.-M.}
  \bibnamefont{Tremblay}}, \bibinfo{journal}{Proceedings of the National
  Academy of Sciences} \textbf{\bibinfo{volume}{118}},
  \bibinfo{pages}{e2106476118} (\bibinfo{year}{2021}),
  \eprint{https://www.pnas.org/doi/pdf/10.1073/pnas.2106476118},
  \urlprefix\url{https://www.pnas.org/doi/abs/10.1073/pnas.2106476118}.

\bibitem[{\citenamefont{O’Mahony et~al.}(2022)\citenamefont{O’Mahony, Ren,
  Chen, Chong, Liu, Eisaki, Uchida, Hamidian, and Davis}}]{omahony2022}
\bibinfo{author}{\bibfnamefont{S.~M.} \bibnamefont{O’Mahony}},
  \bibinfo{author}{\bibfnamefont{W.}~\bibnamefont{Ren}},
  \bibinfo{author}{\bibfnamefont{W.}~\bibnamefont{Chen}},
  \bibinfo{author}{\bibfnamefont{Y.~X.} \bibnamefont{Chong}},
  \bibinfo{author}{\bibfnamefont{X.}~\bibnamefont{Liu}},
  \bibinfo{author}{\bibfnamefont{H.}~\bibnamefont{Eisaki}},
  \bibinfo{author}{\bibfnamefont{S.}~\bibnamefont{Uchida}},
  \bibinfo{author}{\bibfnamefont{M.~H.} \bibnamefont{Hamidian}},
  \bibnamefont{and} \bibinfo{author}{\bibfnamefont{J.~C.~S.}
  \bibnamefont{Davis}}, \bibinfo{journal}{Proceedings of the National Academy
  of Sciences} \textbf{\bibinfo{volume}{119}}, \bibinfo{pages}{e2207449119}
  (\bibinfo{year}{2022}),
  \eprint{https://www.pnas.org/doi/pdf/10.1073/pnas.2207449119},
  \urlprefix\url{https://www.pnas.org/doi/abs/10.1073/pnas.2207449119}.

\bibitem[{\citenamefont{Dean et~al.}(2013{\natexlab{a}})\citenamefont{Dean,
  Dellea, Springell, Yakhou-Harris, Kummer, Brookes, Liu, Sun, Strle, Schmitt
  et~al.}}]{n-mat-dean-lascuo}
\bibinfo{author}{\bibfnamefont{M.~P.~M.} \bibnamefont{Dean}},
  \bibinfo{author}{\bibfnamefont{G.}~\bibnamefont{Dellea}},
  \bibinfo{author}{\bibfnamefont{R.~S.} \bibnamefont{Springell}},
  \bibinfo{author}{\bibfnamefont{F.}~\bibnamefont{Yakhou-Harris}},
  \bibinfo{author}{\bibfnamefont{K.}~\bibnamefont{Kummer}},
  \bibinfo{author}{\bibfnamefont{N.~B.} \bibnamefont{Brookes}},
  \bibinfo{author}{\bibfnamefont{X.}~\bibnamefont{Liu}},
  \bibinfo{author}{\bibfnamefont{Y.-J.} \bibnamefont{Sun}},
  \bibinfo{author}{\bibfnamefont{J.}~\bibnamefont{Strle}},
  \bibinfo{author}{\bibfnamefont{T.}~\bibnamefont{Schmitt}},
  \bibnamefont{et~al.}, \bibinfo{journal}{Nat Mater}
  \textbf{\bibinfo{volume}{12}}, \bibinfo{pages}{1019}
  (\bibinfo{year}{2013}{\natexlab{a}}),
  \urlprefix\url{http://dx.doi.org/10.1038/nmat3723}.

\bibitem[{\citenamefont{Dean et~al.}(2013{\natexlab{b}})\citenamefont{Dean,
  James, Springell, Liu, Monney, Zhou, Konik, Wen, Xu, Gu
  et~al.}}]{dean-bisco-PRL}
\bibinfo{author}{\bibfnamefont{M.~P.~M.} \bibnamefont{Dean}},
  \bibinfo{author}{\bibfnamefont{A.~J.~A.} \bibnamefont{James}},
  \bibinfo{author}{\bibfnamefont{R.~S.} \bibnamefont{Springell}},
  \bibinfo{author}{\bibfnamefont{X.}~\bibnamefont{Liu}},
  \bibinfo{author}{\bibfnamefont{C.}~\bibnamefont{Monney}},
  \bibinfo{author}{\bibfnamefont{K.~J.} \bibnamefont{Zhou}},
  \bibinfo{author}{\bibfnamefont{R.~M.} \bibnamefont{Konik}},
  \bibinfo{author}{\bibfnamefont{J.~S.} \bibnamefont{Wen}},
  \bibinfo{author}{\bibfnamefont{Z.~J.} \bibnamefont{Xu}},
  \bibinfo{author}{\bibfnamefont{G.~D.} \bibnamefont{Gu}},
  \bibnamefont{et~al.}, \bibinfo{journal}{Phys. Rev. Lett.}
  \textbf{\bibinfo{volume}{110}}, \bibinfo{pages}{147001}
  (\bibinfo{year}{2013}{\natexlab{b}}),
  \urlprefix\url{http://link.aps.org/doi/10.1103/PhysRevLett.110.147001}.

\bibitem[{\citenamefont{Guarise et~al.}(2014)\citenamefont{Guarise,
  Dalla~Piazza, Berger, Giannini, Schmitt, R{\o}nnow, Sawatzky, van~den Brink,
  Altenfeld, Eremin et~al.}}]{guarise-bsccoNC}
\bibinfo{author}{\bibfnamefont{M.}~\bibnamefont{Guarise}},
  \bibinfo{author}{\bibfnamefont{B.}~\bibnamefont{Dalla~Piazza}},
  \bibinfo{author}{\bibfnamefont{H.}~\bibnamefont{Berger}},
  \bibinfo{author}{\bibfnamefont{E.}~\bibnamefont{Giannini}},
  \bibinfo{author}{\bibfnamefont{T.}~\bibnamefont{Schmitt}},
  \bibinfo{author}{\bibfnamefont{H.~M.} \bibnamefont{R{\o}nnow}},
  \bibinfo{author}{\bibfnamefont{G.}~\bibnamefont{Sawatzky}},
  \bibinfo{author}{\bibfnamefont{J.}~\bibnamefont{van~den Brink}},
  \bibinfo{author}{\bibfnamefont{D.}~\bibnamefont{Altenfeld}},
  \bibinfo{author}{\bibfnamefont{I.}~\bibnamefont{Eremin}},
  \bibnamefont{et~al.}, \bibinfo{journal}{Nature communications}
  \textbf{\bibinfo{volume}{5}}, \bibinfo{pages}{1} (\bibinfo{year}{2014}).

\bibitem[{\citenamefont{Dean}(2015)}]{Dean20153}
\bibinfo{author}{\bibfnamefont{M.}~\bibnamefont{Dean}},
  \bibinfo{journal}{Journal of Magnetism and Magnetic Materials}
  \textbf{\bibinfo{volume}{376}}, \bibinfo{pages}{3 } (\bibinfo{year}{2015}),
  ISSN \bibinfo{issn}{0304-8853}, \bibinfo{note}{pseudogap, Superconductivity
  and Magnetism},
  \urlprefix\url{http://www.sciencedirect.com/science/article/pii/S0304885314002868}.

\bibitem[{\citenamefont{Peng et~al.}(2018)\citenamefont{Peng, Huang, Fumagalli,
  Minola, Wang, Sun, Ding, Kummer, Zhou, Brookes et~al.}}]{PhysRevB.98.144507}
\bibinfo{author}{\bibfnamefont{Y.~Y.} \bibnamefont{Peng}},
  \bibinfo{author}{\bibfnamefont{E.~W.} \bibnamefont{Huang}},
  \bibinfo{author}{\bibfnamefont{R.}~\bibnamefont{Fumagalli}},
  \bibinfo{author}{\bibfnamefont{M.}~\bibnamefont{Minola}},
  \bibinfo{author}{\bibfnamefont{Y.}~\bibnamefont{Wang}},
  \bibinfo{author}{\bibfnamefont{X.}~\bibnamefont{Sun}},
  \bibinfo{author}{\bibfnamefont{Y.}~\bibnamefont{Ding}},
  \bibinfo{author}{\bibfnamefont{K.}~\bibnamefont{Kummer}},
  \bibinfo{author}{\bibfnamefont{X.~J.} \bibnamefont{Zhou}},
  \bibinfo{author}{\bibfnamefont{N.~B.} \bibnamefont{Brookes}},
  \bibnamefont{et~al.}, \bibinfo{journal}{Phys. Rev. B}
  \textbf{\bibinfo{volume}{98}}, \bibinfo{pages}{144507}
  (\bibinfo{year}{2018}),
  \urlprefix\url{https://link.aps.org/doi/10.1103/PhysRevB.98.144507}.

\bibitem[{\citenamefont{Meyers et~al.}(2017)\citenamefont{Meyers, Miao,
  Walters, Bisogni, Springell, d'Astuto, Dantz, Pelliciari, Huang, Okamoto
  et~al.}}]{meyers-lscoPRB}
\bibinfo{author}{\bibfnamefont{D.}~\bibnamefont{Meyers}},
  \bibinfo{author}{\bibfnamefont{H.}~\bibnamefont{Miao}},
  \bibinfo{author}{\bibfnamefont{A.~C.} \bibnamefont{Walters}},
  \bibinfo{author}{\bibfnamefont{V.}~\bibnamefont{Bisogni}},
  \bibinfo{author}{\bibfnamefont{R.~S.} \bibnamefont{Springell}},
  \bibinfo{author}{\bibfnamefont{M.}~\bibnamefont{d'Astuto}},
  \bibinfo{author}{\bibfnamefont{M.}~\bibnamefont{Dantz}},
  \bibinfo{author}{\bibfnamefont{J.}~\bibnamefont{Pelliciari}},
  \bibinfo{author}{\bibfnamefont{H.~Y.} \bibnamefont{Huang}},
  \bibinfo{author}{\bibfnamefont{J.}~\bibnamefont{Okamoto}},
  \bibnamefont{et~al.}, \bibinfo{journal}{Phys. Rev. B}
  \textbf{\bibinfo{volume}{95}}, \bibinfo{pages}{075139}
  (\bibinfo{year}{2017}),
  \urlprefix\url{https://link.aps.org/doi/10.1103/PhysRevB.95.075139}.

\bibitem[{\citenamefont{Ayral and Parcollet}(2016)}]{Ayral-Parcollet}
\bibinfo{author}{\bibfnamefont{T.}~\bibnamefont{Ayral}} \bibnamefont{and}
  \bibinfo{author}{\bibfnamefont{O.}~\bibnamefont{Parcollet}},
  \bibinfo{journal}{Phys. Rev. B} \textbf{\bibinfo{volume}{93}},
  \bibinfo{pages}{235124} (\bibinfo{year}{2016}),
  \urlprefix\url{https://link.aps.org/doi/10.1103/PhysRevB.93.235124}.

\bibitem[{\citenamefont{Mukuda et~al.}(2012)\citenamefont{Mukuda, Shimizu, Iyo,
  and Kitaoka}}]{mukuda2012}
\bibinfo{author}{\bibfnamefont{H.}~\bibnamefont{Mukuda}},
  \bibinfo{author}{\bibfnamefont{S.}~\bibnamefont{Shimizu}},
  \bibinfo{author}{\bibfnamefont{A.}~\bibnamefont{Iyo}}, \bibnamefont{and}
  \bibinfo{author}{\bibfnamefont{Y.}~\bibnamefont{Kitaoka}},
  \bibinfo{journal}{Journal of the Physical Society of Japan}
  \textbf{\bibinfo{volume}{81}}, \bibinfo{pages}{011008}
  (\bibinfo{year}{2012}), \eprint{https://doi.org/10.1143/JPSJ.81.011008},
  \urlprefix\url{https://doi.org/10.1143/JPSJ.81.011008}.

\bibitem[{\citenamefont{Horn et~al.}(1987)\citenamefont{Horn, Keane, Held,
  Jordan-Sweet, Kaiser, Holtzberg, and Rice}}]{horn1987}
\bibinfo{author}{\bibfnamefont{P.~M.} \bibnamefont{Horn}},
  \bibinfo{author}{\bibfnamefont{D.~T.} \bibnamefont{Keane}},
  \bibinfo{author}{\bibfnamefont{G.~A.} \bibnamefont{Held}},
  \bibinfo{author}{\bibfnamefont{J.~L.} \bibnamefont{Jordan-Sweet}},
  \bibinfo{author}{\bibfnamefont{D.~L.} \bibnamefont{Kaiser}},
  \bibinfo{author}{\bibfnamefont{F.}~\bibnamefont{Holtzberg}},
  \bibnamefont{and} \bibinfo{author}{\bibfnamefont{T.~M.} \bibnamefont{Rice}},
  \bibinfo{journal}{Physical Review Letters} \textbf{\bibinfo{volume}{59}},
  \bibinfo{pages}{2772} (\bibinfo{year}{1987}).

\bibitem[{\citenamefont{Hubbard and Flowers}(1963)}]{hubbard1963}
\bibinfo{author}{\bibfnamefont{J.}~\bibnamefont{Hubbard}} \bibnamefont{and}
  \bibinfo{author}{\bibfnamefont{B.~H.} \bibnamefont{Flowers}},
  \bibinfo{journal}{Proceedings of the Royal Society of London. Series A.
  Mathematical and Physical Sciences} \textbf{\bibinfo{volume}{276}},
  \bibinfo{pages}{238} (\bibinfo{year}{1963}).

\bibitem[{\citenamefont{Gutzwiller}(1963)}]{gutzwiller1963}
\bibinfo{author}{\bibfnamefont{M.~C.} \bibnamefont{Gutzwiller}},
  \bibinfo{journal}{Phys. Rev. Lett.} \textbf{\bibinfo{volume}{10}},
  \bibinfo{pages}{159} (\bibinfo{year}{1963}).

\bibitem[{\citenamefont{Kanamori}(1963)}]{kanamori1963}
\bibinfo{author}{\bibfnamefont{J.}~\bibnamefont{Kanamori}},
  \bibinfo{journal}{Progress of Theoretical Physics}
  \textbf{\bibinfo{volume}{30}}, \bibinfo{pages}{275} (\bibinfo{year}{1963}),
  ISSN \bibinfo{issn}{0033-068X}.

\bibitem[{\citenamefont{Hiroi et~al.}(1994)\citenamefont{Hiroi, Kobayashi, and
  Takano}}]{hiroi}
\bibinfo{author}{\bibfnamefont{Z.}~\bibnamefont{Hiroi}},
  \bibinfo{author}{\bibfnamefont{N.}~\bibnamefont{Kobayashi}},
  \bibnamefont{and} \bibinfo{author}{\bibfnamefont{M.}~\bibnamefont{Takano}},
  \bibinfo{journal}{Nature} \textbf{\bibinfo{volume}{371}},
  \bibinfo{pages}{139} (\bibinfo{year}{1994}).

\bibitem[{\citenamefont{Kohsaka et~al.}(2002)\citenamefont{Kohsaka, Azuma,
  Yamada, Sasagawa, Hanaguri, Takano, and Takagi}}]{kohsaka-jacs}
\bibinfo{author}{\bibfnamefont{Y.}~\bibnamefont{Kohsaka}},
  \bibinfo{author}{\bibfnamefont{M.}~\bibnamefont{Azuma}},
  \bibinfo{author}{\bibfnamefont{I.}~\bibnamefont{Yamada}},
  \bibinfo{author}{\bibfnamefont{T.}~\bibnamefont{Sasagawa}},
  \bibinfo{author}{\bibfnamefont{T.}~\bibnamefont{Hanaguri}},
  \bibinfo{author}{\bibfnamefont{M.}~\bibnamefont{Takano}}, \bibnamefont{and}
  \bibinfo{author}{\bibfnamefont{H.}~\bibnamefont{Takagi}},
  \bibinfo{journal}{Journal of the American Chemical Society}
  \textbf{\bibinfo{volume}{124}}, \bibinfo{pages}{12275}
  (\bibinfo{year}{2002}), \bibinfo{note}{pMID: 12371870},
  \eprint{http://dx.doi.org/10.1021/ja026680i},
  \urlprefix\url{http://dx.doi.org/10.1021/ja026680i}.

\bibitem[{\citenamefont{Baptiste et~al.}(2018)\citenamefont{Baptiste, Azuma,
  Yu, Giura, and d'Astuto}}]{baptiste-ccoc-cif}
\bibinfo{author}{\bibfnamefont{B.}~\bibnamefont{Baptiste}},
  \bibinfo{author}{\bibfnamefont{M.}~\bibnamefont{Azuma}},
  \bibinfo{author}{\bibfnamefont{R.}~\bibnamefont{Yu}},
  \bibinfo{author}{\bibfnamefont{P.}~\bibnamefont{Giura}}, \bibnamefont{and}
  \bibinfo{author}{\bibfnamefont{M.}~\bibnamefont{d'Astuto}},
  \bibinfo{journal}{IUCrData} \textbf{\bibinfo{volume}{3}},
  \bibinfo{pages}{x181645} (\bibinfo{year}{2018}),
  \urlprefix\url{https://doi.org/10.1107/S2414314618016450}.

\bibitem[{\citenamefont{Vaknin et~al.}(1997)\citenamefont{Vaknin, Miller, and
  Zarestky}}]{vaknin_prb}
\bibinfo{author}{\bibfnamefont{D.}~\bibnamefont{Vaknin}},
  \bibinfo{author}{\bibfnamefont{L.~L.} \bibnamefont{Miller}},
  \bibnamefont{and} \bibinfo{author}{\bibfnamefont{J.~L.}
  \bibnamefont{Zarestky}}, \bibinfo{journal}{Phys. Rev. B}
  \textbf{\bibinfo{volume}{56}}, \bibinfo{pages}{8351} (\bibinfo{year}{1997}).

\bibitem[{\citenamefont{Yamada et~al.}(2005)\citenamefont{Yamada, Belik, Azuma,
  Harjo, Kamiyama, Shimakawa, and Takano}}]{Yamada2005}
\bibinfo{author}{\bibfnamefont{I.}~\bibnamefont{Yamada}},
  \bibinfo{author}{\bibfnamefont{A.~A.} \bibnamefont{Belik}},
  \bibinfo{author}{\bibfnamefont{M.}~\bibnamefont{Azuma}},
  \bibinfo{author}{\bibfnamefont{S.}~\bibnamefont{Harjo}},
  \bibinfo{author}{\bibfnamefont{T.}~\bibnamefont{Kamiyama}},
  \bibinfo{author}{\bibfnamefont{Y.}~\bibnamefont{Shimakawa}},
  \bibnamefont{and} \bibinfo{author}{\bibfnamefont{M.}~\bibnamefont{Takano}},
  \bibinfo{journal}{Phys. Rev. B} \textbf{\bibinfo{volume}{72}},
  \bibinfo{pages}{224503} (\bibinfo{year}{2005}),
  \urlprefix\url{https://link.aps.org/doi/10.1103/PhysRevB.72.224503}.

\bibitem[{\citenamefont{Hirayama et~al.}(2018)\citenamefont{Hirayama, Yamaji,
  Misawa, and Imada}}]{hirayama2018}
\bibinfo{author}{\bibfnamefont{M.}~\bibnamefont{Hirayama}},
  \bibinfo{author}{\bibfnamefont{Y.}~\bibnamefont{Yamaji}},
  \bibinfo{author}{\bibfnamefont{T.}~\bibnamefont{Misawa}}, \bibnamefont{and}
  \bibinfo{author}{\bibfnamefont{M.}~\bibnamefont{Imada}},
  \bibinfo{journal}{Physical Review B} \textbf{\bibinfo{volume}{98}},
  \bibinfo{pages}{134501} (\bibinfo{year}{2018}).

\bibitem[{\citenamefont{Mor{\'e}e et~al.}(2022)\citenamefont{Mor{\'e}e,
  Hirayama, Schmid, Yamaji, and Imada}}]{moree2022}
\bibinfo{author}{\bibfnamefont{J.-B.} \bibnamefont{Mor{\'e}e}},
  \bibinfo{author}{\bibfnamefont{M.}~\bibnamefont{Hirayama}},
  \bibinfo{author}{\bibfnamefont{M.~T.} \bibnamefont{Schmid}},
  \bibinfo{author}{\bibfnamefont{Y.}~\bibnamefont{Yamaji}}, \bibnamefont{and}
  \bibinfo{author}{\bibfnamefont{M.}~\bibnamefont{Imada}},
  \bibinfo{journal}{Physical Review B} \textbf{\bibinfo{volume}{106}},
  \bibinfo{pages}{235150} (\bibinfo{year}{2022}).

\bibitem[{\citenamefont{Dalla~Piazza et~al.}(2012)\citenamefont{Dalla~Piazza,
  Mourigal, Guarise, Berger, Schmitt, Zhou, Grioni, and
  R\o{}nnow}}]{dalla_piazza_rapidB}
\bibinfo{author}{\bibfnamefont{B.}~\bibnamefont{Dalla~Piazza}},
  \bibinfo{author}{\bibfnamefont{M.}~\bibnamefont{Mourigal}},
  \bibinfo{author}{\bibfnamefont{M.}~\bibnamefont{Guarise}},
  \bibinfo{author}{\bibfnamefont{H.}~\bibnamefont{Berger}},
  \bibinfo{author}{\bibfnamefont{T.}~\bibnamefont{Schmitt}},
  \bibinfo{author}{\bibfnamefont{K.~J.} \bibnamefont{Zhou}},
  \bibinfo{author}{\bibfnamefont{M.}~\bibnamefont{Grioni}}, \bibnamefont{and}
  \bibinfo{author}{\bibfnamefont{H.~M.} \bibnamefont{R\o{}nnow}},
  \bibinfo{journal}{Phys. Rev. B} \textbf{\bibinfo{volume}{85}},
  \bibinfo{pages}{100508(R)} (\bibinfo{year}{2012}),
  \urlprefix\url{http://link.aps.org/doi/10.1103/PhysRevB.85.100508}.

\bibitem[{\citenamefont{Kim et~al.}(1998)\citenamefont{Kim, White, Shen,
  Tohyama, Shibata, Maekawa, Wells, Kim, Birgeneau, and
  Kastner}}]{PhysRevLett.80.4245}
\bibinfo{author}{\bibfnamefont{C.}~\bibnamefont{Kim}},
  \bibinfo{author}{\bibfnamefont{P.~J.} \bibnamefont{White}},
  \bibinfo{author}{\bibfnamefont{Z.-X.} \bibnamefont{Shen}},
  \bibinfo{author}{\bibfnamefont{T.}~\bibnamefont{Tohyama}},
  \bibinfo{author}{\bibfnamefont{Y.}~\bibnamefont{Shibata}},
  \bibinfo{author}{\bibfnamefont{S.}~\bibnamefont{Maekawa}},
  \bibinfo{author}{\bibfnamefont{B.~O.} \bibnamefont{Wells}},
  \bibinfo{author}{\bibfnamefont{Y.~J.} \bibnamefont{Kim}},
  \bibinfo{author}{\bibfnamefont{R.~J.} \bibnamefont{Birgeneau}},
  \bibnamefont{and} \bibinfo{author}{\bibfnamefont{M.~A.}
  \bibnamefont{Kastner}}, \bibinfo{journal}{Phys. Rev. Lett.}
  \textbf{\bibinfo{volume}{80}}, \bibinfo{pages}{4245} (\bibinfo{year}{1998}),
  \urlprefix\url{https://link.aps.org/doi/10.1103/PhysRevLett.80.4245}.

\bibitem[{\citenamefont{Lebert et~al.}(2017)\citenamefont{Lebert, Dean,
  Nicolaou, Pelliciari, Dantz, Schmitt, Yu, Azuma, Castellan, Miao
  et~al.}}]{ccoc-sw-blair1}
\bibinfo{author}{\bibfnamefont{B.~W.} \bibnamefont{Lebert}},
  \bibinfo{author}{\bibfnamefont{M.~P.~M.} \bibnamefont{Dean}},
  \bibinfo{author}{\bibfnamefont{A.}~\bibnamefont{Nicolaou}},
  \bibinfo{author}{\bibfnamefont{J.}~\bibnamefont{Pelliciari}},
  \bibinfo{author}{\bibfnamefont{M.}~\bibnamefont{Dantz}},
  \bibinfo{author}{\bibfnamefont{T.}~\bibnamefont{Schmitt}},
  \bibinfo{author}{\bibfnamefont{R.}~\bibnamefont{Yu}},
  \bibinfo{author}{\bibfnamefont{M.}~\bibnamefont{Azuma}},
  \bibinfo{author}{\bibfnamefont{J.-P.} \bibnamefont{Castellan}},
  \bibinfo{author}{\bibfnamefont{H.}~\bibnamefont{Miao}}, \bibnamefont{et~al.},
  \bibinfo{journal}{Phys. Rev. B} \textbf{\bibinfo{volume}{95}},
  \bibinfo{pages}{155110} (\bibinfo{year}{2017}),
  \urlprefix\url{https://link.aps.org/doi/10.1103/PhysRevB.95.155110}.

\bibitem[{\citenamefont{Jia et~al.}(2014)\citenamefont{Jia, Nowadnick,
  Wohlfeld, Kung, Chen, Johnston, Tohyama, Moritz, and Devereaux}}]{Jia2014}
\bibinfo{author}{\bibfnamefont{C.~J.} \bibnamefont{Jia}},
  \bibinfo{author}{\bibfnamefont{E.~A.} \bibnamefont{Nowadnick}},
  \bibinfo{author}{\bibfnamefont{K.}~\bibnamefont{Wohlfeld}},
  \bibinfo{author}{\bibfnamefont{Y.~F.} \bibnamefont{Kung}},
  \bibinfo{author}{\bibfnamefont{C.-C.} \bibnamefont{Chen}},
  \bibinfo{author}{\bibfnamefont{S.}~\bibnamefont{Johnston}},
  \bibinfo{author}{\bibfnamefont{T.}~\bibnamefont{Tohyama}},
  \bibinfo{author}{\bibfnamefont{B.}~\bibnamefont{Moritz}}, \bibnamefont{and}
  \bibinfo{author}{\bibfnamefont{T.~P.} \bibnamefont{Devereaux}},
  \bibinfo{journal}{Nature Communications} \textbf{\bibinfo{volume}{5}},
  \bibinfo{pages}{3314} (\bibinfo{year}{2014}), ISSN \bibinfo{issn}{2041-1723},
  \urlprefix\url{https://doi.org/10.1038/ncomms4314}.

\bibitem[{\citenamefont{Chaix et~al.}(2018)\citenamefont{Chaix, Huang, Gerber,
  Lu, Jia, Huang, McNally, Wang, Vernay, Keren et~al.}}]{chaix2018}
\bibinfo{author}{\bibfnamefont{L.}~\bibnamefont{Chaix}},
  \bibinfo{author}{\bibfnamefont{E.~W.} \bibnamefont{Huang}},
  \bibinfo{author}{\bibfnamefont{S.}~\bibnamefont{Gerber}},
  \bibinfo{author}{\bibfnamefont{X.}~\bibnamefont{Lu}},
  \bibinfo{author}{\bibfnamefont{C.}~\bibnamefont{Jia}},
  \bibinfo{author}{\bibfnamefont{Y.}~\bibnamefont{Huang}},
  \bibinfo{author}{\bibfnamefont{D.~E.} \bibnamefont{McNally}},
  \bibinfo{author}{\bibfnamefont{Y.}~\bibnamefont{Wang}},
  \bibinfo{author}{\bibfnamefont{F.~H.} \bibnamefont{Vernay}},
  \bibinfo{author}{\bibfnamefont{A.}~\bibnamefont{Keren}},
  \bibnamefont{et~al.}, \bibinfo{journal}{Phys. Rev. B}
  \textbf{\bibinfo{volume}{97}}, \bibinfo{pages}{155144}
  (\bibinfo{year}{2018}).

\bibitem[{\citenamefont{Brookes et~al.}(2018)\citenamefont{Brookes,
  Yakhou-Harris, Kummer, Fondacaro, Cezar, Betto, Velez-Fort, Amorese,
  Ghiringhelli, Braicovich et~al.}}]{Brookes2018}
\bibinfo{author}{\bibfnamefont{N.}~\bibnamefont{Brookes}},
  \bibinfo{author}{\bibfnamefont{F.}~\bibnamefont{Yakhou-Harris}},
  \bibinfo{author}{\bibfnamefont{K.}~\bibnamefont{Kummer}},
  \bibinfo{author}{\bibfnamefont{A.}~\bibnamefont{Fondacaro}},
  \bibinfo{author}{\bibfnamefont{J.}~\bibnamefont{Cezar}},
  \bibinfo{author}{\bibfnamefont{D.}~\bibnamefont{Betto}},
  \bibinfo{author}{\bibfnamefont{E.}~\bibnamefont{Velez-Fort}},
  \bibinfo{author}{\bibfnamefont{A.}~\bibnamefont{Amorese}},
  \bibinfo{author}{\bibfnamefont{G.}~\bibnamefont{Ghiringhelli}},
  \bibinfo{author}{\bibfnamefont{L.}~\bibnamefont{Braicovich}},
  \bibnamefont{et~al.}, \bibinfo{journal}{Nuclear Instruments and Methods in
  Physics Research Section A: Accelerators, Spectrometers, Detectors and
  Associated Equipment} \textbf{\bibinfo{volume}{903}}, \bibinfo{pages}{175}
  (\bibinfo{year}{2018}), ISSN \bibinfo{issn}{0168-9002},
  \urlprefix\url{https://www.sciencedirect.com/science/article/pii/S0168900218308234}.

\bibitem[{\citenamefont{Braicovich
  et~al.}(2010{\natexlab{a}})\citenamefont{Braicovich, Moretti~Sala, Ament,
  Bisogni, Minola, Balestrino, Di~Castro, De~Luca, Salluzzo, Ghiringhelli
  et~al.}}]{pol-dep-prb}
\bibinfo{author}{\bibfnamefont{L.}~\bibnamefont{Braicovich}},
  \bibinfo{author}{\bibfnamefont{M.}~\bibnamefont{Moretti~Sala}},
  \bibinfo{author}{\bibfnamefont{L.~J.~P.} \bibnamefont{Ament}},
  \bibinfo{author}{\bibfnamefont{V.}~\bibnamefont{Bisogni}},
  \bibinfo{author}{\bibfnamefont{M.}~\bibnamefont{Minola}},
  \bibinfo{author}{\bibfnamefont{G.}~\bibnamefont{Balestrino}},
  \bibinfo{author}{\bibfnamefont{D.}~\bibnamefont{Di~Castro}},
  \bibinfo{author}{\bibfnamefont{G.~M.} \bibnamefont{De~Luca}},
  \bibinfo{author}{\bibfnamefont{M.}~\bibnamefont{Salluzzo}},
  \bibinfo{author}{\bibfnamefont{G.}~\bibnamefont{Ghiringhelli}},
  \bibnamefont{et~al.}, \bibinfo{journal}{Phys. Rev. B}
  \textbf{\bibinfo{volume}{81}}, \bibinfo{pages}{174533}
  (\bibinfo{year}{2010}{\natexlab{a}}).

\bibitem[{\citenamefont{Civelli et~al.}(2005)\citenamefont{Civelli, Capone,
  Kancharla, Parcollet, and Kotliar}}]{civelli2005}
\bibinfo{author}{\bibfnamefont{M.}~\bibnamefont{Civelli}},
  \bibinfo{author}{\bibfnamefont{M.}~\bibnamefont{Capone}},
  \bibinfo{author}{\bibfnamefont{S.~S.} \bibnamefont{Kancharla}},
  \bibinfo{author}{\bibfnamefont{O.}~\bibnamefont{Parcollet}},
  \bibnamefont{and} \bibinfo{author}{\bibfnamefont{G.}~\bibnamefont{Kotliar}},
  \bibinfo{journal}{Physical review letters} \textbf{\bibinfo{volume}{95}},
  \bibinfo{pages}{106402} (\bibinfo{year}{2005}).

\bibitem[{\citenamefont{Kyung et~al.}(2006)\citenamefont{Kyung, Kancharla,
  S\'en\'echal, Tremblay, Civelli, and Kotliar}}]{kyung2006}
\bibinfo{author}{\bibfnamefont{B.}~\bibnamefont{Kyung}},
  \bibinfo{author}{\bibfnamefont{S.~S.} \bibnamefont{Kancharla}},
  \bibinfo{author}{\bibfnamefont{D.}~\bibnamefont{S\'en\'echal}},
  \bibinfo{author}{\bibfnamefont{A.-M.~S.} \bibnamefont{Tremblay}},
  \bibinfo{author}{\bibfnamefont{M.}~\bibnamefont{Civelli}}, \bibnamefont{and}
  \bibinfo{author}{\bibfnamefont{G.}~\bibnamefont{Kotliar}},
  \bibinfo{journal}{Physical Review B} \textbf{\bibinfo{volume}{73}},
  \bibinfo{pages}{165114} (\bibinfo{year}{2006}).

\bibitem[{\citenamefont{Macridin et~al.}(2006)\citenamefont{Macridin, Jarrell,
  Maier, Kent, and D'Azevedo}}]{macridin2006}
\bibinfo{author}{\bibfnamefont{A.}~\bibnamefont{Macridin}},
  \bibinfo{author}{\bibfnamefont{M.}~\bibnamefont{Jarrell}},
  \bibinfo{author}{\bibfnamefont{T.}~\bibnamefont{Maier}},
  \bibinfo{author}{\bibfnamefont{P.~R.~C.} \bibnamefont{Kent}},
  \bibnamefont{and}
  \bibinfo{author}{\bibfnamefont{E.}~\bibnamefont{D'Azevedo}},
  \bibinfo{journal}{Physical Review Letters} \textbf{\bibinfo{volume}{97}},
  \bibinfo{pages}{036401} (\bibinfo{year}{2006}).

\bibitem[{\citenamefont{Damascelli et~al.}(2003)\citenamefont{Damascelli,
  Hussain, and Shen}}]{damascelli2003}
\bibinfo{author}{\bibfnamefont{A.}~\bibnamefont{Damascelli}},
  \bibinfo{author}{\bibfnamefont{Z.}~\bibnamefont{Hussain}}, \bibnamefont{and}
  \bibinfo{author}{\bibfnamefont{Z.-X.} \bibnamefont{Shen}},
  \bibinfo{journal}{Reviews of Modern Physics} \textbf{\bibinfo{volume}{75}},
  \bibinfo{pages}{473} (\bibinfo{year}{2003}).

\bibitem[{\citenamefont{Ronning et~al.}(2005)\citenamefont{Ronning, Shen,
  Armitage, Damascelli, Lu, Shen, Miller, and Kim}}]{ronning2005}
\bibinfo{author}{\bibfnamefont{F.}~\bibnamefont{Ronning}},
  \bibinfo{author}{\bibfnamefont{K.~M.} \bibnamefont{Shen}},
  \bibinfo{author}{\bibfnamefont{N.~P.} \bibnamefont{Armitage}},
  \bibinfo{author}{\bibfnamefont{A.}~\bibnamefont{Damascelli}},
  \bibinfo{author}{\bibfnamefont{D.~H.} \bibnamefont{Lu}},
  \bibinfo{author}{\bibfnamefont{Z.-X.} \bibnamefont{Shen}},
  \bibinfo{author}{\bibfnamefont{L.~L.} \bibnamefont{Miller}},
  \bibnamefont{and} \bibinfo{author}{\bibfnamefont{C.}~\bibnamefont{Kim}},
  \bibinfo{journal}{Physical Review B} \textbf{\bibinfo{volume}{71}},
  \bibinfo{pages}{094518} (\bibinfo{year}{2005}).

\bibitem[{\citenamefont{Macridin et~al.}(2007)\citenamefont{Macridin, Jarrell,
  Maier, and Scalapino}}]{macridin2007}
\bibinfo{author}{\bibfnamefont{A.}~\bibnamefont{Macridin}},
  \bibinfo{author}{\bibfnamefont{M.}~\bibnamefont{Jarrell}},
  \bibinfo{author}{\bibfnamefont{T.}~\bibnamefont{Maier}}, \bibnamefont{and}
  \bibinfo{author}{\bibfnamefont{D.~J.} \bibnamefont{Scalapino}},
  \bibinfo{journal}{Physical Review Letters} \textbf{\bibinfo{volume}{99}},
  \bibinfo{pages}{237001} (\bibinfo{year}{2007}).

\bibitem[{\citenamefont{Manousakis}(2007)}]{manousakis2007}
\bibinfo{author}{\bibfnamefont{E.}~\bibnamefont{Manousakis}},
  \bibinfo{journal}{Physical Review B} \textbf{\bibinfo{volume}{75}},
  \bibinfo{pages}{035106} (\bibinfo{year}{2007}).

\bibitem[{\citenamefont{Wang et~al.}(2015)\citenamefont{Wang, Wohlfeld, Moritz,
  Jia, van Veenendaal, Wu, Chen, and Devereaux}}]{wang2015}
\bibinfo{author}{\bibfnamefont{Y.}~\bibnamefont{Wang}},
  \bibinfo{author}{\bibfnamefont{K.}~\bibnamefont{Wohlfeld}},
  \bibinfo{author}{\bibfnamefont{B.}~\bibnamefont{Moritz}},
  \bibinfo{author}{\bibfnamefont{C.~J.} \bibnamefont{Jia}},
  \bibinfo{author}{\bibfnamefont{M.}~\bibnamefont{van Veenendaal}},
  \bibinfo{author}{\bibfnamefont{K.}~\bibnamefont{Wu}},
  \bibinfo{author}{\bibfnamefont{C.-C.} \bibnamefont{Chen}}, \bibnamefont{and}
  \bibinfo{author}{\bibfnamefont{T.~P.} \bibnamefont{Devereaux}},
  \bibinfo{journal}{Physical Review B} \textbf{\bibinfo{volume}{92}},
  \bibinfo{pages}{075119} (\bibinfo{year}{2015}).

\bibitem[{\citenamefont{Stepanov et~al.}(2018)\citenamefont{Stepanov, Peters,
  Krivenko, Lichtenstein, Katsnelson, and Rubtsov}}]{stepanov2018}
\bibinfo{author}{\bibfnamefont{E.~A.} \bibnamefont{Stepanov}},
  \bibinfo{author}{\bibfnamefont{L.}~\bibnamefont{Peters}},
  \bibinfo{author}{\bibfnamefont{I.~S.} \bibnamefont{Krivenko}},
  \bibinfo{author}{\bibfnamefont{A.~I.} \bibnamefont{Lichtenstein}},
  \bibinfo{author}{\bibfnamefont{M.~I.} \bibnamefont{Katsnelson}},
  \bibnamefont{and} \bibinfo{author}{\bibfnamefont{A.~N.}
  \bibnamefont{Rubtsov}}, \bibinfo{journal}{npj Quantum Materials}
  \textbf{\bibinfo{volume}{3}}, \bibinfo{pages}{54} (\bibinfo{year}{2018}).

\bibitem[{\citenamefont{Blaha et~al.}(2021)\citenamefont{Blaha, Schwarz,
  K.~H.~Madsen, Kvasnicka, Luitz, Laskowski, Tran, and D.~Marks}}]{wien2k}
\bibinfo{author}{\bibfnamefont{P.}~\bibnamefont{Blaha}},
  \bibinfo{author}{\bibfnamefont{K.}~\bibnamefont{Schwarz}},
  \bibinfo{author}{\bibfnamefont{G.}~\bibnamefont{K.~H.~Madsen}},
  \bibinfo{author}{\bibfnamefont{D.}~\bibnamefont{Kvasnicka}},
  \bibinfo{author}{\bibfnamefont{J.}~\bibnamefont{Luitz}},
  \bibinfo{author}{\bibfnamefont{R.}~\bibnamefont{Laskowski}},
  \bibinfo{author}{\bibfnamefont{F.}~\bibnamefont{Tran}}, \bibnamefont{and}
  \bibinfo{author}{\bibfnamefont{L.}~\bibnamefont{D.~Marks}},
  \emph{\bibinfo{title}{An {{Augmented Plane Wave}} + {{Local Orbitals
  Program}} for {{Calculating Crystal Properties}}}}
  (\bibinfo{publisher}{Karlheinz Schwarz, Techn. Universität Wien, Austria},
  \bibinfo{year}{2021}), \bibinfo{edition}{revised edition wien2k\_21.1} ed.,
  ISBN \bibinfo{isbn}{3-9501031-1-2}.

\bibitem[{\citenamefont{Blaha et~al.}(2020)\citenamefont{Blaha, Schwarz, Tran,
  Laskowski, Madsen, and Marks}}]{blaha2020}
\bibinfo{author}{\bibfnamefont{P.}~\bibnamefont{Blaha}},
  \bibinfo{author}{\bibfnamefont{K.}~\bibnamefont{Schwarz}},
  \bibinfo{author}{\bibfnamefont{F.}~\bibnamefont{Tran}},
  \bibinfo{author}{\bibfnamefont{R.}~\bibnamefont{Laskowski}},
  \bibinfo{author}{\bibfnamefont{G.~K.~H.} \bibnamefont{Madsen}},
  \bibnamefont{and} \bibinfo{author}{\bibfnamefont{L.~D.} \bibnamefont{Marks}},
  \bibinfo{journal}{J. Chem. Phys.} \textbf{\bibinfo{volume}{152}},
  \bibinfo{pages}{074101} (\bibinfo{year}{2020}), ISSN
  \bibinfo{issn}{0021-9606, 1089-7690}.

\bibitem[{\citenamefont{Marzari and Vanderbilt}(1997)}]{marzari1997}
\bibinfo{author}{\bibfnamefont{N.}~\bibnamefont{Marzari}} \bibnamefont{and}
  \bibinfo{author}{\bibfnamefont{D.}~\bibnamefont{Vanderbilt}},
  \bibinfo{journal}{Phys. Rev. B} \textbf{\bibinfo{volume}{56}},
  \bibinfo{pages}{12847} (\bibinfo{year}{1997}).

\bibitem[{\citenamefont{Souza et~al.}(2001)\citenamefont{Souza, Marzari, and
  Vanderbilt}}]{souza2001}
\bibinfo{author}{\bibfnamefont{I.}~\bibnamefont{Souza}},
  \bibinfo{author}{\bibfnamefont{N.}~\bibnamefont{Marzari}}, \bibnamefont{and}
  \bibinfo{author}{\bibfnamefont{D.}~\bibnamefont{Vanderbilt}},
  \bibinfo{journal}{Phys. Rev. B} \textbf{\bibinfo{volume}{65}},
  \bibinfo{pages}{035109} (\bibinfo{year}{2001}).

\bibitem[{\citenamefont{Mostofi et~al.}(2008)\citenamefont{Mostofi, Yates, Lee,
  Souza, Vanderbilt, and Marzari}}]{mostofi2008}
\bibinfo{author}{\bibfnamefont{A.~A.} \bibnamefont{Mostofi}},
  \bibinfo{author}{\bibfnamefont{J.~R.} \bibnamefont{Yates}},
  \bibinfo{author}{\bibfnamefont{Y.-S.} \bibnamefont{Lee}},
  \bibinfo{author}{\bibfnamefont{I.}~\bibnamefont{Souza}},
  \bibinfo{author}{\bibfnamefont{D.}~\bibnamefont{Vanderbilt}},
  \bibnamefont{and} \bibinfo{author}{\bibfnamefont{N.}~\bibnamefont{Marzari}},
  \bibinfo{journal}{Computer Physics Communications}
  \textbf{\bibinfo{volume}{178}}, \bibinfo{pages}{685} (\bibinfo{year}{2008}),
  ISSN \bibinfo{issn}{0010-4655}.

\bibitem[{\citenamefont{Mostofi et~al.}(2014)\citenamefont{Mostofi, Yates,
  Pizzi, Lee, Souza, Vanderbilt, and Marzari}}]{mostofi2014}
\bibinfo{author}{\bibfnamefont{A.~A.} \bibnamefont{Mostofi}},
  \bibinfo{author}{\bibfnamefont{J.~R.} \bibnamefont{Yates}},
  \bibinfo{author}{\bibfnamefont{G.}~\bibnamefont{Pizzi}},
  \bibinfo{author}{\bibfnamefont{Y.-S.} \bibnamefont{Lee}},
  \bibinfo{author}{\bibfnamefont{I.}~\bibnamefont{Souza}},
  \bibinfo{author}{\bibfnamefont{D.}~\bibnamefont{Vanderbilt}},
  \bibnamefont{and} \bibinfo{author}{\bibfnamefont{N.}~\bibnamefont{Marzari}},
  \bibinfo{journal}{Computer Physics Communications}
  \textbf{\bibinfo{volume}{185}}, \bibinfo{pages}{2309} (\bibinfo{year}{2014}),
  ISSN \bibinfo{issn}{0010-4655}.

\bibitem[{\citenamefont{Lichtenstein and Katsnelson}(1998)}]{lichtenstein1998}
\bibinfo{author}{\bibfnamefont{A.~I.} \bibnamefont{Lichtenstein}}
  \bibnamefont{and} \bibinfo{author}{\bibfnamefont{M.~I.}
  \bibnamefont{Katsnelson}}, \bibinfo{journal}{Phys. Rev. B}
  \textbf{\bibinfo{volume}{57}}, \bibinfo{pages}{6884} (\bibinfo{year}{1998}).

\bibitem[{\citenamefont{Lichtenstein and Katsnelson}(2000)}]{lichtenstein2000}
\bibinfo{author}{\bibfnamefont{A.~I.} \bibnamefont{Lichtenstein}}
  \bibnamefont{and} \bibinfo{author}{\bibfnamefont{M.~I.}
  \bibnamefont{Katsnelson}}, \bibinfo{journal}{Phys. Rev. B}
  \textbf{\bibinfo{volume}{62}}, \bibinfo{pages}{R9283} (\bibinfo{year}{2000}).

\bibitem[{\citenamefont{Kotliar et~al.}(2001)\citenamefont{Kotliar, Savrasov,
  P{\'a}lsson, and Biroli}}]{kotliar2001}
\bibinfo{author}{\bibfnamefont{G.}~\bibnamefont{Kotliar}},
  \bibinfo{author}{\bibfnamefont{S.~Y.} \bibnamefont{Savrasov}},
  \bibinfo{author}{\bibfnamefont{G.}~\bibnamefont{P{\'a}lsson}},
  \bibnamefont{and} \bibinfo{author}{\bibfnamefont{G.}~\bibnamefont{Biroli}},
  \bibinfo{journal}{Phys. Rev. Lett.} \textbf{\bibinfo{volume}{87}},
  \bibinfo{pages}{186401} (\bibinfo{year}{2001}).

\bibitem[{\citenamefont{Maier et~al.}(2005)\citenamefont{Maier, Jarrell,
  Pruschke, and Hettler}}]{maier2005}
\bibinfo{author}{\bibfnamefont{T.}~\bibnamefont{Maier}},
  \bibinfo{author}{\bibfnamefont{M.}~\bibnamefont{Jarrell}},
  \bibinfo{author}{\bibfnamefont{T.}~\bibnamefont{Pruschke}}, \bibnamefont{and}
  \bibinfo{author}{\bibfnamefont{M.~H.} \bibnamefont{Hettler}},
  \bibinfo{journal}{Rev. Mod. Phys.} \textbf{\bibinfo{volume}{77}},
  \bibinfo{pages}{1027} (\bibinfo{year}{2005}).

\bibitem[{\citenamefont{Georges et~al.}(1996)\citenamefont{Georges, Kotliar,
  Krauth, and Rozenberg}}]{georges1996}
\bibinfo{author}{\bibfnamefont{A.}~\bibnamefont{Georges}},
  \bibinfo{author}{\bibfnamefont{G.}~\bibnamefont{Kotliar}},
  \bibinfo{author}{\bibfnamefont{W.}~\bibnamefont{Krauth}}, \bibnamefont{and}
  \bibinfo{author}{\bibfnamefont{M.~J.} \bibnamefont{Rozenberg}},
  \bibinfo{journal}{Rev. Mod. Phys.} \textbf{\bibinfo{volume}{68}},
  \bibinfo{pages}{13} (\bibinfo{year}{1996}).

\bibitem[{\citenamefont{Rubtsov et~al.}(2005)\citenamefont{Rubtsov, Savkin, and
  Lichtenstein}}]{rubtsov2005}
\bibinfo{author}{\bibfnamefont{A.~N.} \bibnamefont{Rubtsov}},
  \bibinfo{author}{\bibfnamefont{V.~V.} \bibnamefont{Savkin}},
  \bibnamefont{and} \bibinfo{author}{\bibfnamefont{A.~I.}
  \bibnamefont{Lichtenstein}}, \bibinfo{journal}{Phys. Rev. B}
  \textbf{\bibinfo{volume}{72}}, \bibinfo{pages}{035122}
  (\bibinfo{year}{2005}), ISSN \bibinfo{issn}{1098-0121, 1550-235X}.

\bibitem[{\citenamefont{Shinaoka et~al.}(2014)\citenamefont{Shinaoka, Dolfi,
  Troyer, and Werner}}]{shinaoka2014}
\bibinfo{author}{\bibfnamefont{H.}~\bibnamefont{Shinaoka}},
  \bibinfo{author}{\bibfnamefont{M.}~\bibnamefont{Dolfi}},
  \bibinfo{author}{\bibfnamefont{M.}~\bibnamefont{Troyer}}, \bibnamefont{and}
  \bibinfo{author}{\bibfnamefont{P.}~\bibnamefont{Werner}},
  \bibinfo{journal}{J. Stat. Mech.} \textbf{\bibinfo{volume}{2014}},
  \bibinfo{pages}{P06012} (\bibinfo{year}{2014}), ISSN
  \bibinfo{issn}{1742-5468}.

\bibitem[{\citenamefont{Gaenko et~al.}(2017)\citenamefont{Gaenko, Antipov,
  Carcassi, Chen, Chen, Dong, Gamper, Gukelberger, Igarashi, Iskakov
  et~al.}}]{gaenko2017}
\bibinfo{author}{\bibfnamefont{A.}~\bibnamefont{Gaenko}},
  \bibinfo{author}{\bibfnamefont{A.~E.} \bibnamefont{Antipov}},
  \bibinfo{author}{\bibfnamefont{G.}~\bibnamefont{Carcassi}},
  \bibinfo{author}{\bibfnamefont{T.}~\bibnamefont{Chen}},
  \bibinfo{author}{\bibfnamefont{X.}~\bibnamefont{Chen}},
  \bibinfo{author}{\bibfnamefont{Q.}~\bibnamefont{Dong}},
  \bibinfo{author}{\bibfnamefont{L.}~\bibnamefont{Gamper}},
  \bibinfo{author}{\bibfnamefont{J.}~\bibnamefont{Gukelberger}},
  \bibinfo{author}{\bibfnamefont{R.}~\bibnamefont{Igarashi}},
  \bibinfo{author}{\bibfnamefont{S.}~\bibnamefont{Iskakov}},
  \bibnamefont{et~al.}, \bibinfo{journal}{Computer Physics Communications}
  \textbf{\bibinfo{volume}{213}}, \bibinfo{pages}{235} (\bibinfo{year}{2017}).

\bibitem[{\citenamefont{Levy et~al.}(2017)\citenamefont{Levy, LeBlanc, and
  Gull}}]{levy2017}
\bibinfo{author}{\bibfnamefont{R.}~\bibnamefont{Levy}},
  \bibinfo{author}{\bibfnamefont{J.}~\bibnamefont{LeBlanc}}, \bibnamefont{and}
  \bibinfo{author}{\bibfnamefont{E.}~\bibnamefont{Gull}},
  \bibinfo{journal}{Computer Physics Communications}
  \textbf{\bibinfo{volume}{215}}, \bibinfo{pages}{149} (\bibinfo{year}{2017}).

\bibitem[{\citenamefont{Fumagalli et~al.}(2019)\citenamefont{Fumagalli,
  Braicovich, Minola, Peng, Kummer, Betto, Rossi,
  Lefran\ifmmode~\mbox{\c{c}}\else \c{c}\fi{}ois, Morawe, Salluzzo
  et~al.}}]{PhysRevB.99.134517}
\bibinfo{author}{\bibfnamefont{R.}~\bibnamefont{Fumagalli}},
  \bibinfo{author}{\bibfnamefont{L.}~\bibnamefont{Braicovich}},
  \bibinfo{author}{\bibfnamefont{M.}~\bibnamefont{Minola}},
  \bibinfo{author}{\bibfnamefont{Y.~Y.} \bibnamefont{Peng}},
  \bibinfo{author}{\bibfnamefont{K.}~\bibnamefont{Kummer}},
  \bibinfo{author}{\bibfnamefont{D.}~\bibnamefont{Betto}},
  \bibinfo{author}{\bibfnamefont{M.}~\bibnamefont{Rossi}},
  \bibinfo{author}{\bibfnamefont{E.}~\bibnamefont{Lefran\ifmmode~\mbox{\c{c}}\else
  \c{c}\fi{}ois}}, \bibinfo{author}{\bibfnamefont{C.}~\bibnamefont{Morawe}},
  \bibinfo{author}{\bibfnamefont{M.}~\bibnamefont{Salluzzo}},
  \bibnamefont{et~al.}, \bibinfo{journal}{Phys. Rev. B}
  \textbf{\bibinfo{volume}{99}}, \bibinfo{pages}{134517}
  (\bibinfo{year}{2019}),
  \urlprefix\url{https://link.aps.org/doi/10.1103/PhysRevB.99.134517}.

\bibitem[{\citenamefont{Braicovich
  et~al.}(2010{\natexlab{b}})\citenamefont{Braicovich, van~den Brink, Bisogni,
  Moretti~Sala, Ament, Brookes, De~Luca, Salluzzo, Schmitt, Strocov
  et~al.}}]{ghiringhelli-prl-rixs-mag}
\bibinfo{author}{\bibfnamefont{L.}~\bibnamefont{Braicovich}},
  \bibinfo{author}{\bibfnamefont{J.}~\bibnamefont{van~den Brink}},
  \bibinfo{author}{\bibfnamefont{V.}~\bibnamefont{Bisogni}},
  \bibinfo{author}{\bibfnamefont{M.}~\bibnamefont{Moretti~Sala}},
  \bibinfo{author}{\bibfnamefont{L.~J.~P.} \bibnamefont{Ament}},
  \bibinfo{author}{\bibfnamefont{N.~B.} \bibnamefont{Brookes}},
  \bibinfo{author}{\bibfnamefont{G.~M.} \bibnamefont{De~Luca}},
  \bibinfo{author}{\bibfnamefont{M.}~\bibnamefont{Salluzzo}},
  \bibinfo{author}{\bibfnamefont{T.}~\bibnamefont{Schmitt}},
  \bibinfo{author}{\bibfnamefont{V.~N.} \bibnamefont{Strocov}},
  \bibnamefont{et~al.}, \bibinfo{journal}{Phys. Rev. Lett.}
  \textbf{\bibinfo{volume}{104}}, \bibinfo{pages}{077002}
  (\bibinfo{year}{2010}{\natexlab{b}}),
  \urlprefix\url{http://link.aps.org/doi/10.1103/PhysRevLett.104.077002}.

\bibitem[{\citenamefont{Dean et~al.}(2012)\citenamefont{Dean, Springell,
  Monney, Zhou, Pereiro, Božović, Dalla~Piazza, Rønnow, Morenzoni, van~den
  Brink et~al.}}]{Dean2012}
\bibinfo{author}{\bibfnamefont{M.~P.~M.} \bibnamefont{Dean}},
  \bibinfo{author}{\bibfnamefont{R.~S.} \bibnamefont{Springell}},
  \bibinfo{author}{\bibfnamefont{C.}~\bibnamefont{Monney}},
  \bibinfo{author}{\bibfnamefont{K.~J.} \bibnamefont{Zhou}},
  \bibinfo{author}{\bibfnamefont{J.}~\bibnamefont{Pereiro}},
  \bibinfo{author}{\bibfnamefont{I.}~\bibnamefont{Božović}},
  \bibinfo{author}{\bibfnamefont{B.}~\bibnamefont{Dalla~Piazza}},
  \bibinfo{author}{\bibfnamefont{H.~M.} \bibnamefont{Rønnow}},
  \bibinfo{author}{\bibfnamefont{E.}~\bibnamefont{Morenzoni}},
  \bibinfo{author}{\bibfnamefont{J.}~\bibnamefont{van~den Brink}},
  \bibnamefont{et~al.}, \bibinfo{journal}{Nature Materials}
  \textbf{\bibinfo{volume}{11}}, \bibinfo{pages}{850} (\bibinfo{year}{2012}),
  ISSN \bibinfo{issn}{1476-4660},
  \urlprefix\url{https://doi.org/10.1038/nmat3409}.

\bibitem[{\citenamefont{{Moretti Sala} et~al.}(2011)\citenamefont{{Moretti
  Sala}, Bisogni, Aruta, Balestrino, Berger, Brookes, {De Luca}, {Di Castro},
  Grioni, Guarise et~al.}}]{MorettiSala2011}
\bibinfo{author}{\bibfnamefont{M.}~\bibnamefont{{Moretti Sala}}},
  \bibinfo{author}{\bibfnamefont{V.}~\bibnamefont{Bisogni}},
  \bibinfo{author}{\bibfnamefont{C.}~\bibnamefont{Aruta}},
  \bibinfo{author}{\bibfnamefont{G.}~\bibnamefont{Balestrino}},
  \bibinfo{author}{\bibfnamefont{H.}~\bibnamefont{Berger}},
  \bibinfo{author}{\bibfnamefont{N.~B.} \bibnamefont{Brookes}},
  \bibinfo{author}{\bibfnamefont{G.~M.} \bibnamefont{{De Luca}}},
  \bibinfo{author}{\bibfnamefont{D.}~\bibnamefont{{Di Castro}}},
  \bibinfo{author}{\bibfnamefont{M.}~\bibnamefont{Grioni}},
  \bibinfo{author}{\bibfnamefont{M.}~\bibnamefont{Guarise}},
  \bibnamefont{et~al.}, \bibinfo{journal}{New Journal of Physics}
  \textbf{\bibinfo{volume}{13}} (\bibinfo{year}{2011}), ISSN
  \bibinfo{issn}{13672630}, \eprint{1009.4882}.

\bibitem[{\citenamefont{Minola et~al.}(2015)\citenamefont{Minola, Dellea,
  Gretarsson, Peng, Lu, Porras, Loew, Yakhou, Brookes, Huang
  et~al.}}]{PhysRevLett.114.217003}
\bibinfo{author}{\bibfnamefont{M.}~\bibnamefont{Minola}},
  \bibinfo{author}{\bibfnamefont{G.}~\bibnamefont{Dellea}},
  \bibinfo{author}{\bibfnamefont{H.}~\bibnamefont{Gretarsson}},
  \bibinfo{author}{\bibfnamefont{Y.~Y.} \bibnamefont{Peng}},
  \bibinfo{author}{\bibfnamefont{Y.}~\bibnamefont{Lu}},
  \bibinfo{author}{\bibfnamefont{J.}~\bibnamefont{Porras}},
  \bibinfo{author}{\bibfnamefont{T.}~\bibnamefont{Loew}},
  \bibinfo{author}{\bibfnamefont{F.}~\bibnamefont{Yakhou}},
  \bibinfo{author}{\bibfnamefont{N.~B.} \bibnamefont{Brookes}},
  \bibinfo{author}{\bibfnamefont{Y.~B.} \bibnamefont{Huang}},
  \bibnamefont{et~al.}, \bibinfo{journal}{Phys. Rev. Lett.}
  \textbf{\bibinfo{volume}{114}}, \bibinfo{pages}{217003}
  (\bibinfo{year}{2015}),
  \urlprefix\url{https://link.aps.org/doi/10.1103/PhysRevLett.114.217003}.

\bibitem[{\citenamefont{Rossi et~al.}(2019)\citenamefont{Rossi, Arpaia,
  Fumagalli, Moretti~Sala, Betto, Kummer, De~Luca, van~den Brink, Salluzzo,
  Brookes et~al.}}]{PhysRevLett.123.027001}
\bibinfo{author}{\bibfnamefont{M.}~\bibnamefont{Rossi}},
  \bibinfo{author}{\bibfnamefont{R.}~\bibnamefont{Arpaia}},
  \bibinfo{author}{\bibfnamefont{R.}~\bibnamefont{Fumagalli}},
  \bibinfo{author}{\bibfnamefont{M.}~\bibnamefont{Moretti~Sala}},
  \bibinfo{author}{\bibfnamefont{D.}~\bibnamefont{Betto}},
  \bibinfo{author}{\bibfnamefont{K.}~\bibnamefont{Kummer}},
  \bibinfo{author}{\bibfnamefont{G.~M.} \bibnamefont{De~Luca}},
  \bibinfo{author}{\bibfnamefont{J.}~\bibnamefont{van~den Brink}},
  \bibinfo{author}{\bibfnamefont{M.}~\bibnamefont{Salluzzo}},
  \bibinfo{author}{\bibfnamefont{N.~B.} \bibnamefont{Brookes}},
  \bibnamefont{et~al.}, \bibinfo{journal}{Phys. Rev. Lett.}
  \textbf{\bibinfo{volume}{123}}, \bibinfo{pages}{027001}
  (\bibinfo{year}{2019}),
  \urlprefix\url{https://link.aps.org/doi/10.1103/PhysRevLett.123.027001}.

\bibitem[{\citenamefont{Lamsal and Montfrooij}(2016)}]{PhysRevB.93.214513}
\bibinfo{author}{\bibfnamefont{J.}~\bibnamefont{Lamsal}} \bibnamefont{and}
  \bibinfo{author}{\bibfnamefont{W.}~\bibnamefont{Montfrooij}},
  \bibinfo{journal}{Phys. Rev. B} \textbf{\bibinfo{volume}{93}},
  \bibinfo{pages}{214513} (\bibinfo{year}{2016}),
  \urlprefix\url{https://link.aps.org/doi/10.1103/PhysRevB.93.214513}.

\bibitem[{\citenamefont{Ohishi et~al.}(2005)\citenamefont{Ohishi, Yamada, Koda,
  Higemoto, R.~Saha, Kadono, M.~Kojima, Azuma, and Takano}}]{ohishi2005}
\bibinfo{author}{\bibfnamefont{K.}~\bibnamefont{Ohishi}},
  \bibinfo{author}{\bibfnamefont{I.}~\bibnamefont{Yamada}},
  \bibinfo{author}{\bibfnamefont{A.}~\bibnamefont{Koda}},
  \bibinfo{author}{\bibfnamefont{W.}~\bibnamefont{Higemoto}},
  \bibinfo{author}{\bibfnamefont{S.}~\bibnamefont{R.~Saha}},
  \bibinfo{author}{\bibfnamefont{R.}~\bibnamefont{Kadono}},
  \bibinfo{author}{\bibfnamefont{K.}~\bibnamefont{M.~Kojima}},
  \bibinfo{author}{\bibfnamefont{M.}~\bibnamefont{Azuma}}, \bibnamefont{and}
  \bibinfo{author}{\bibfnamefont{M.}~\bibnamefont{Takano}},
  \bibinfo{journal}{Journal of the Physical Society of Japan}
  \textbf{\bibinfo{volume}{74}}, \bibinfo{pages}{2408} (\bibinfo{year}{2005}).

\bibitem[{\citenamefont{Coldea et~al.}(2001)\citenamefont{Coldea, Hayden,
  Aeppli, Perring, Frost, Mason, Cheong, and Fisk}}]{coldea-prl-lco}
\bibinfo{author}{\bibfnamefont{R.}~\bibnamefont{Coldea}},
  \bibinfo{author}{\bibfnamefont{S.~M.} \bibnamefont{Hayden}},
  \bibinfo{author}{\bibfnamefont{G.}~\bibnamefont{Aeppli}},
  \bibinfo{author}{\bibfnamefont{T.~G.} \bibnamefont{Perring}},
  \bibinfo{author}{\bibfnamefont{C.~D.} \bibnamefont{Frost}},
  \bibinfo{author}{\bibfnamefont{T.~E.} \bibnamefont{Mason}},
  \bibinfo{author}{\bibfnamefont{S.-W.} \bibnamefont{Cheong}},
  \bibnamefont{and} \bibinfo{author}{\bibfnamefont{Z.}~\bibnamefont{Fisk}},
  \bibinfo{journal}{Phys. Rev. Lett.} \textbf{\bibinfo{volume}{86}},
  \bibinfo{pages}{5377} (\bibinfo{year}{2001}),
  \urlprefix\url{http://link.aps.org/doi/10.1103/PhysRevLett.86.5377}.

\bibitem[{\citenamefont{Delannoy et~al.}(2009)\citenamefont{Delannoy, Gingras,
  Holdsworth, and Tremblay}}]{delannoy2009}
\bibinfo{author}{\bibfnamefont{J.-Y.~P.} \bibnamefont{Delannoy}},
  \bibinfo{author}{\bibfnamefont{M.~J.~P.} \bibnamefont{Gingras}},
  \bibinfo{author}{\bibfnamefont{P.~C.~W.} \bibnamefont{Holdsworth}},
  \bibnamefont{and} \bibinfo{author}{\bibfnamefont{A.-M.~S.}
  \bibnamefont{Tremblay}}, \bibinfo{journal}{Phys. Rev. B}
  \textbf{\bibinfo{volume}{79}}, \bibinfo{pages}{235130}
  (\bibinfo{year}{2009}), ISSN \bibinfo{issn}{1098-0121, 1550-235X}.

\bibitem[{\citenamefont{Peng et~al.}(2017)\citenamefont{Peng, Dellea, Minola,
  Conni, Amorese, Di~Castro, De~Luca, Kummer, Salluzzo, Sun et~al.}}]{peng2017}
\bibinfo{author}{\bibfnamefont{Y.~Y.} \bibnamefont{Peng}},
  \bibinfo{author}{\bibfnamefont{G.}~\bibnamefont{Dellea}},
  \bibinfo{author}{\bibfnamefont{M.}~\bibnamefont{Minola}},
  \bibinfo{author}{\bibfnamefont{M.}~\bibnamefont{Conni}},
  \bibinfo{author}{\bibfnamefont{A.}~\bibnamefont{Amorese}},
  \bibinfo{author}{\bibfnamefont{D.}~\bibnamefont{Di~Castro}},
  \bibinfo{author}{\bibfnamefont{G.~M.} \bibnamefont{De~Luca}},
  \bibinfo{author}{\bibfnamefont{K.}~\bibnamefont{Kummer}},
  \bibinfo{author}{\bibfnamefont{M.}~\bibnamefont{Salluzzo}},
  \bibinfo{author}{\bibfnamefont{X.}~\bibnamefont{Sun}}, \bibnamefont{et~al.},
  \bibinfo{journal}{Nature Physics} \textbf{\bibinfo{volume}{13}},
  \bibinfo{pages}{1201} (\bibinfo{year}{2017}), ISSN \bibinfo{issn}{1745-2481},
  \urlprefix\url{https://doi.org/10.1038/nphys4248}.

\bibitem[{\citenamefont{Betto et~al.}(2021)\citenamefont{Betto, Fumagalli,
  Martinelli, Rossi, Piombo, Yoshimi, Di~Castro, Di~Gennaro, Sambri, Bonn
  et~al.}}]{PhysRevB.103.L140409}
\bibinfo{author}{\bibfnamefont{D.}~\bibnamefont{Betto}},
  \bibinfo{author}{\bibfnamefont{R.}~\bibnamefont{Fumagalli}},
  \bibinfo{author}{\bibfnamefont{L.}~\bibnamefont{Martinelli}},
  \bibinfo{author}{\bibfnamefont{M.}~\bibnamefont{Rossi}},
  \bibinfo{author}{\bibfnamefont{R.}~\bibnamefont{Piombo}},
  \bibinfo{author}{\bibfnamefont{K.}~\bibnamefont{Yoshimi}},
  \bibinfo{author}{\bibfnamefont{D.}~\bibnamefont{Di~Castro}},
  \bibinfo{author}{\bibfnamefont{E.}~\bibnamefont{Di~Gennaro}},
  \bibinfo{author}{\bibfnamefont{A.}~\bibnamefont{Sambri}},
  \bibinfo{author}{\bibfnamefont{D.}~\bibnamefont{Bonn}}, \bibnamefont{et~al.},
  \bibinfo{journal}{Phys. Rev. B} \textbf{\bibinfo{volume}{103}},
  \bibinfo{pages}{L140409} (\bibinfo{year}{2021}),
  \urlprefix\url{https://link.aps.org/doi/10.1103/PhysRevB.103.L140409}.

\bibitem[{\citenamefont{Guarise et~al.}(2010)\citenamefont{Guarise,
  Dalla~Piazza, Moretti~Sala, Ghiringhelli, Braicovich, Berger, Hancock,
  van~der Marel, Schmitt, Strocov et~al.}}]{guarise-rixs-mag-prl}
\bibinfo{author}{\bibfnamefont{M.}~\bibnamefont{Guarise}},
  \bibinfo{author}{\bibfnamefont{B.}~\bibnamefont{Dalla~Piazza}},
  \bibinfo{author}{\bibfnamefont{M.}~\bibnamefont{Moretti~Sala}},
  \bibinfo{author}{\bibfnamefont{G.}~\bibnamefont{Ghiringhelli}},
  \bibinfo{author}{\bibfnamefont{L.}~\bibnamefont{Braicovich}},
  \bibinfo{author}{\bibfnamefont{H.}~\bibnamefont{Berger}},
  \bibinfo{author}{\bibfnamefont{J.~N.} \bibnamefont{Hancock}},
  \bibinfo{author}{\bibfnamefont{D.}~\bibnamefont{van~der Marel}},
  \bibinfo{author}{\bibfnamefont{T.}~\bibnamefont{Schmitt}},
  \bibinfo{author}{\bibfnamefont{V.~N.} \bibnamefont{Strocov}},
  \bibnamefont{et~al.}, \bibinfo{journal}{Phys. Rev. Lett.}
  \textbf{\bibinfo{volume}{105}}, \bibinfo{pages}{157006}
  (\bibinfo{year}{2010}),
  \urlprefix\url{http://link.aps.org/doi/10.1103/PhysRevLett.105.157006}.

\bibitem[{\citenamefont{Aryasetiawan et~al.}(2004)\citenamefont{Aryasetiawan,
  Imada, Georges, Kotliar, Biermann, and Lichtenstein}}]{Aryasetiawan2004}
\bibinfo{author}{\bibfnamefont{F.}~\bibnamefont{Aryasetiawan}},
  \bibinfo{author}{\bibfnamefont{M.}~\bibnamefont{Imada}},
  \bibinfo{author}{\bibfnamefont{A.}~\bibnamefont{Georges}},
  \bibinfo{author}{\bibfnamefont{G.}~\bibnamefont{Kotliar}},
  \bibinfo{author}{\bibfnamefont{S.}~\bibnamefont{Biermann}}, \bibnamefont{and}
  \bibinfo{author}{\bibfnamefont{A.~I.} \bibnamefont{Lichtenstein}},
  \bibinfo{journal}{Phys. Rev. B} \textbf{\bibinfo{volume}{70}},
  \bibinfo{pages}{195104} (\bibinfo{year}{2004}),
  \urlprefix\url{https://link.aps.org/doi/10.1103/PhysRevB.70.195104}.

\bibitem[{\citenamefont{Headings et~al.}(2010)\citenamefont{Headings, Hayden,
  Coldea, and Perring}}]{Headings2010}
\bibinfo{author}{\bibfnamefont{N.~S.} \bibnamefont{Headings}},
  \bibinfo{author}{\bibfnamefont{S.~M.} \bibnamefont{Hayden}},
  \bibinfo{author}{\bibfnamefont{R.}~\bibnamefont{Coldea}}, \bibnamefont{and}
  \bibinfo{author}{\bibfnamefont{T.~G.} \bibnamefont{Perring}},
  \bibinfo{journal}{Physical Review Letters} \textbf{\bibinfo{volume}{105}},
  \bibinfo{pages}{247001} (\bibinfo{year}{2010}), ISSN
  \bibinfo{issn}{0031-9007},
  \urlprefix\url{http://link.aps.org/doi/10.1103/PhysRevLett.105.247001}.

\bibitem[{\citenamefont{Ivashko et~al.}(2019)\citenamefont{Ivashko, Horio, Wan,
  Christensen, McNally, Paris, Tseng, Shaik, R{\o}nnow, Wei
  et~al.}}]{ivashko2019}
\bibinfo{author}{\bibfnamefont{O.}~\bibnamefont{Ivashko}},
  \bibinfo{author}{\bibfnamefont{M.}~\bibnamefont{Horio}},
  \bibinfo{author}{\bibfnamefont{W.}~\bibnamefont{Wan}},
  \bibinfo{author}{\bibfnamefont{N.}~\bibnamefont{Christensen}},
  \bibinfo{author}{\bibfnamefont{D.}~\bibnamefont{McNally}},
  \bibinfo{author}{\bibfnamefont{E.}~\bibnamefont{Paris}},
  \bibinfo{author}{\bibfnamefont{Y.}~\bibnamefont{Tseng}},
  \bibinfo{author}{\bibfnamefont{N.}~\bibnamefont{Shaik}},
  \bibinfo{author}{\bibfnamefont{H.}~\bibnamefont{R{\o}nnow}},
  \bibinfo{author}{\bibfnamefont{H.}~\bibnamefont{Wei}}, \bibnamefont{et~al.},
  \bibinfo{journal}{Nature communications} \textbf{\bibinfo{volume}{10}},
  \bibinfo{pages}{786} (\bibinfo{year}{2019}).

\bibitem[{\citenamefont{Dean et~al.}(2014)\citenamefont{Dean, James, Walters,
  Bisogni, Jarrige, H\"ucker, Giannini, Fujita, Pelliciari, Huang
  et~al.}}]{PhysRevB.90.220506}
\bibinfo{author}{\bibfnamefont{M.~P.~M.} \bibnamefont{Dean}},
  \bibinfo{author}{\bibfnamefont{A.~J.~A.} \bibnamefont{James}},
  \bibinfo{author}{\bibfnamefont{A.~C.} \bibnamefont{Walters}},
  \bibinfo{author}{\bibfnamefont{V.}~\bibnamefont{Bisogni}},
  \bibinfo{author}{\bibfnamefont{I.}~\bibnamefont{Jarrige}},
  \bibinfo{author}{\bibfnamefont{M.}~\bibnamefont{H\"ucker}},
  \bibinfo{author}{\bibfnamefont{E.}~\bibnamefont{Giannini}},
  \bibinfo{author}{\bibfnamefont{M.}~\bibnamefont{Fujita}},
  \bibinfo{author}{\bibfnamefont{J.}~\bibnamefont{Pelliciari}},
  \bibinfo{author}{\bibfnamefont{Y.~B.} \bibnamefont{Huang}},
  \bibnamefont{et~al.}, \bibinfo{journal}{Phys. Rev. B}
  \textbf{\bibinfo{volume}{90}}, \bibinfo{pages}{220506(R)}
  (\bibinfo{year}{2014}),
  \urlprefix\url{https://link.aps.org/doi/10.1103/PhysRevB.90.220506}.

\bibitem[{\citenamefont{Ellis et~al.}(2015)\citenamefont{Ellis, Huang,
  Olalde-Velasco, Dantz, Pelliciari, Drachuck, Ofer, Bazalitsky, Berger,
  Schmitt et~al.}}]{PhysRevB.92.104507}
\bibinfo{author}{\bibfnamefont{D.~S.} \bibnamefont{Ellis}},
  \bibinfo{author}{\bibfnamefont{Y.-B.} \bibnamefont{Huang}},
  \bibinfo{author}{\bibfnamefont{P.}~\bibnamefont{Olalde-Velasco}},
  \bibinfo{author}{\bibfnamefont{M.}~\bibnamefont{Dantz}},
  \bibinfo{author}{\bibfnamefont{J.}~\bibnamefont{Pelliciari}},
  \bibinfo{author}{\bibfnamefont{G.}~\bibnamefont{Drachuck}},
  \bibinfo{author}{\bibfnamefont{R.}~\bibnamefont{Ofer}},
  \bibinfo{author}{\bibfnamefont{G.}~\bibnamefont{Bazalitsky}},
  \bibinfo{author}{\bibfnamefont{J.}~\bibnamefont{Berger}},
  \bibinfo{author}{\bibfnamefont{T.}~\bibnamefont{Schmitt}},
  \bibnamefont{et~al.}, \bibinfo{journal}{Phys. Rev. B}
  \textbf{\bibinfo{volume}{92}}, \bibinfo{pages}{104507}
  (\bibinfo{year}{2015}),
  \urlprefix\url{https://link.aps.org/doi/10.1103/PhysRevB.92.104507}.

\end{thebibliography}
\end{document}